\documentclass[%
 aip,
 amsmath,amssymb,
 reprint,%
]{revtex4-1}

\usepackage{graphicx}% Include figure files
\usepackage{dcolumn}% Align table columns on decimal point
\usepackage{bm}% bold math
\usepackage{braket}

\usepackage[utf8]{inputenc}
\usepackage{bm}
\usepackage{mathptmx}
\usepackage{amsmath}
\usepackage{amssymb}
\usepackage{etoolbox}

\makeatletter
\def\@email#1#2{%
 \endgroup
 \patchcmd{\titleblock@produce}
  {\frontmatter@RRAPformat}
  {\frontmatter@RRAPformat{\produce@RRAP{*#1\href{mailto:#2}{#2}}}\frontmatter@RRAPformat}
  {}{}
}%
\makeatother
\begin{document}

\preprint{AIP/123-QED}

\title[Voltage Dependent Symmetry Breaking in Spin-$1/2$ Paramagnetic Molecular Junctions Driven Out of Equilibrium]{Voltage Dependent Symmetry Breaking in Spin-$1/2$ Paramagnetic Molecular Junctions Driven Out of Equilibrium}
% Force line breaks with \\
\author{Maria Fernanda Bustos Velasquez}% 
\affiliation{Departamento de Física y Geología, Universidad de Pamplona. Pamplona, Colombia}
 \email{juan.vasquez@unipamplona.edu.co}
\author{Juan Camilo Velez Quinones}%
 \affiliation{Departamento de Física, Universidad del Valle, A.A. 25360, Cali, Colombia}

\author{Karem Cecilia Rodriguez Ramirez}
\affiliation{Departamento de Física, Universidad del Valle, A.A. 25360, Cali, Colombia}
\author{Juan David Vasquez Jaramillo}
\affiliation{Departamento de Física y Geología, Universidad de Pamplona. Pamplona, Colombia}
\homepage{http://https://www.unipamplona.edu.co/fisica/}
\affiliation{Departamento de Física, Universidad del Valle, A.A. 25360, Cali, Colombia}%%Lines break automatically or can be forced with \\
\date{\today}% 
\begin{abstract}
In the present work, we consider a spin 1/2 paramagnetic dimer embedded in a magnetic tunnel junction driven out of equilibrium by means of an applied voltage and an applied temperature gradient, in the presence of an external magnetic bias, which can be turn on/off. Here, we derive the spin excitation spectrum for a two spin 1/2 system analitically and show that an external magnetic field is required to lift the degeneracy in the triplet state, whether both spin units experience the magnetic field in the same direction or in opposite staggered directions. We show that an applied bias voltage in the absence of a magnetic field, transfers the magnetization from the magnetic leads into the spin units embedded in the molecule, hence breaking a symmetry and lifting the triplet degeneracy in the absence of external magnetization. From this theoretical demonstration, we then propose two schemes of magnetic driving to lift spin degenracy in the spin excitation spectrum, which we named PMD (Parallel Magnetic Driving) and AMD (Anti-Parallel Magnetic Driving). We argue that a symmetry breaking is then required to project the spin configuration of a dimer into quantum transport measurements, and hence propose different schemes to detect the spin triplet state and the spin singlet state using differential conductivity measurements in the non-degenerate non-symmetric configuration of the spin dimer, and compare our results with the ones reported in the literature.
\end{abstract}

\maketitle

\section{Introduction}
Single and few magnetic atoms on surface have been investigated in recent years as a versatile physical system for tailoring nanomagnets through engineering symmetric \cite{Wortmann2001,Urdampilleta2011,Khajetoorians2012,Urdampilleta2013,Wagner2013a,Spinelli2014} and chiral anti-symmetric exchange interactions \cite{Khajetoorians2016,Steinbrecher2018}, for engineering and customizing single ion magnetic anisotropy \cite{Heinrich2015,Jacobson2015,Meng2016,VasquezJaramillo2018a}, improving spin filtering effects \cite{Aharony2011,Matityahu2013} and realizing magnetically doped topological insulators \cite{Mi2013,Loptien2018}, among others, typically, within experimental contexts related to scanning tunneling microscopy and spectroscopy \cite{Wortmann2001,Ternes2015}, inelastic spin flip spectroscopy \cite{Chen2008} and electron spin resonance spectroscopy \cite{Fransson2008a,Misochko2012,Wu2017}. From the experimental ground, it has been demonstrated that, by atomic scale resolution positioning of individual magnetic dopants in metallic thin films, different types of exchange interactions can be engineered \cite{Khajetoorians2012}, for instance, symmetric and isotropic exchange interactions (RKKY like), chiral anti-symmetric exchange interactions (Dzyaloshinskii-Moriya like) \cite{Khajetoorians2016}, and by engineering the adsorption site, single ion anisotropy can be engineered \cite{Baumann2015}, as well as by the application of an electric field using a scanning tunneling probe \cite{Loth2012}. For each different type of magnetic configuration engineered in the reported experiments by J. Wiebe \textit{et.al} \cite{Khajetoorians2012,Khajetoorians2016,Steinbrecher2018}, the single atom magnetometry has been performed as a function of an applied external magnetic fields and the corresponding conductance measurement has been done using scanning tunneling spectroscopy (STS), showing clear signatures of the magnetization of the cluster of adsorbed magnetic atoms on surface. From the theoretical perspective, there are two dominant approaches reported in literature to determine the spin excitation spectrum, hence the molecule magnetization and its effect on the single magnetic atom conductivity in scanning tunneling spectroscopy set ups. First, J. Fransson proposed a theory for spin polarized scanning tunneling spectroscopy on spins adsorbed on surface and spin inelastic tunneling spectroscopy on a single spin adsorbed on surface \cite{Fransson2010}, showing the signatures of the spin excitation spectrum in the differential conductivity. Moreover, the same author studied the itinerant electron mediated dynamical exchange interaction between individual impurities \cite{Fransson2010dyn} extending the previous work to several spin units, from where it was found that its components decay similarly to the decay trend present in the RKKY interaction, which was generalized in \cite{Fransson2014,Fransson2017} to include Dzyaloshinskii-Moriya like and symmetric anisotropic contributions. Both of these approaches for equilibrium and nonequilibrium conditions, hence, defining a voltage and temperature dependent effective spin Hamiltonian, from where the nonequilibrium excitation spectrum can be obtained and hence, its effect on quantum transport as investigated in \cite{Hammar2016,Vasquez-Jaramillo2017,VasquezJaramillo2018b,Hammar2018}. Secondly, the work by Markus Ternes makes use of third order perturbation theory to determine the spin excitation spectrum \cite{Ternes2015} with the differentiating factor that, the spin Hamiltonian is independent of voltage and temperature in contrast to J. Fransson's approach where the exchange interactions that define the Hamiltonian are tuned through a nonequilibrium drive whether voltage bias or temperature gradient. While Ternes approach has been found to be very precise in reproducing the experimental trends in differential conductivity experiments performed in the context of scanning tunneling spectroscopy \cite{Khajetoorians2016,Steinbrecher2018}, Fransson's approach is ideal to determine the effect of other degrees of freedom such as vibrations on the exchange interactions and anisotropies \cite{VasquezJaramillo2018a} and hence in the way the latter affects the quantum transport. 
%%%%%%%%%%%%%%%%%%%%%%%%%%%%%%%%%%%%%%%%%%%%%%%%%%%%%%%%%%%%%%%%%%%%%%%%%%%%%%%%%%%%%%%%%%%%%%%%%%%%%%%%%%%%%%%%%%%%%%%%%%%%%%%%%%%%%%%%%%%%%%%%%%%%%%%%%%%%%%%%%%%%%%%%%%%%%%%%%%%%%%%%%%%%%%%%%%%%%%%%%%%%%%%%%%%%%%%%%%%%%%%%%%%%%%%%%%%%%%%%%%%%%%%%%%%%%%%%%%%%%%%%%%%%%%%%%%%%%%%%%%%%%%%%%%%%%%%%%%%%%%%%%%%%%%%%%%%%%%%%%%%%%%%%%%%%%%%%%%%%%%%%%%%%%%%%%%%%%%%%%%%%%%%%%%%%%%%%%%%%%%%%%%%%%%%%%%%%%%%%
\section{Contribution of This Paper}
In this work we consider a spin 1/2 paramagnetic dimer, placed in between two ferromagnetic leads in a spin molecular junction contribution and we study the molecular magnetization, spin-spin interactions and its effect on quantum transport. Here, we show analitically the impossibility of molecular moment formation in the absence of an external magnetic field or transferral of magnetization by proximity from ferromagnetic leads, which is a fundamental part of spin selective quantum transport in molecular junctions with spin degrees of freedom, in contrast to what is argued in\cite{Vasquez-Jaramillo2017}. Moreover, here we show that in the absence of external magnetic fields, it is enough to project the magnetization of ferromagnetic leads by proximity into the molecule using a bias voltage in the junction, hence, \textsl{inducing a voltage dependent symmetry breaking in the molecule}. Finally, we propose two schemes for driving magnetically the molecule with external fields, one being parallel magnetic driving (PMD) and the other one being anti-parallel magnetic driving (AMD), which in turn have different effects of the RKKY-Like interaction between spin units in the dimer and hence have different electric differential conductivity responses.
%%%%%%%%%%%%%%%%%%%%%%%%%%%%%%%%%%%%%%%%%%%%%%%%%%%%%%%%%%%%%%%%%%%%%%%%%%%%%%%%%%%%%%%%%%%%%%%%%%%%%%%%%%%%%%%%%%%%%%%%%%%%%%%%%%%%%%%%%%%%%%%%%%%%%%%%%%%%%%%%%%%%%%%%%%%%%%%%%%%%%%%%%%%%%%%%%%%%%%%%%%%%%%%%%%%%%%%%%%%%%%%%%%%%%%%%%%%%%%%%%%%%%%%%%%%%%%%%%%%%%%%%%%%%%%%%%%%%%%%%%%%%%%%%%%%%%%%%%%%%%%%%%%%%%%%%%%%%%%%%%%%%%%%%%%%%%%%%%%%%%%%%%%%%%%%%%%%%%%%%%%%%%%%%%%%%%%%%%%%%%%%%%%%%%%%%%%%%%%%%
\section{Methodology and Mathematical Models}
In previous theoretical work by H. Hammar and J. Fransson \textsl{et.al} a paramagnetic coupled spin pair was shown to switch its magnetization as a function of an applied bias voltage and hence it exhibited voltage induced switching dynamics in the I-V curve, and in a complementary theoretical work by J.D Vasquez Jaramillo and J. Fransson, we showed that this voltage induced switching dynamics produced a commutation between zero, low and high thermal conductivity regimes through a metallic junction with the former paramagnetic coupled spin pair embedded within it, and hence producing different regimes of Peltier heating shown in the evaluation of the nonequilibrium Seebeck coefficient as a function of temperature gradient across the junction and as a function of applied bias voltage, this, in addition to the proposal of an analogous Seebeck coefficient associated with the heat current. This work considered a metallic tunnel junction with a magnetic molecule consisting of two spin units embedded within it. Both metallic contacts can be either normal of ferromagnetic metals, and the magnetic molecule couples through tunneling to both of these metals. The model Hamiltonian for the system of study shown in figure \ref{dimer} is given by:

\begin{figure*}
\includegraphics[width=1\textwidth]{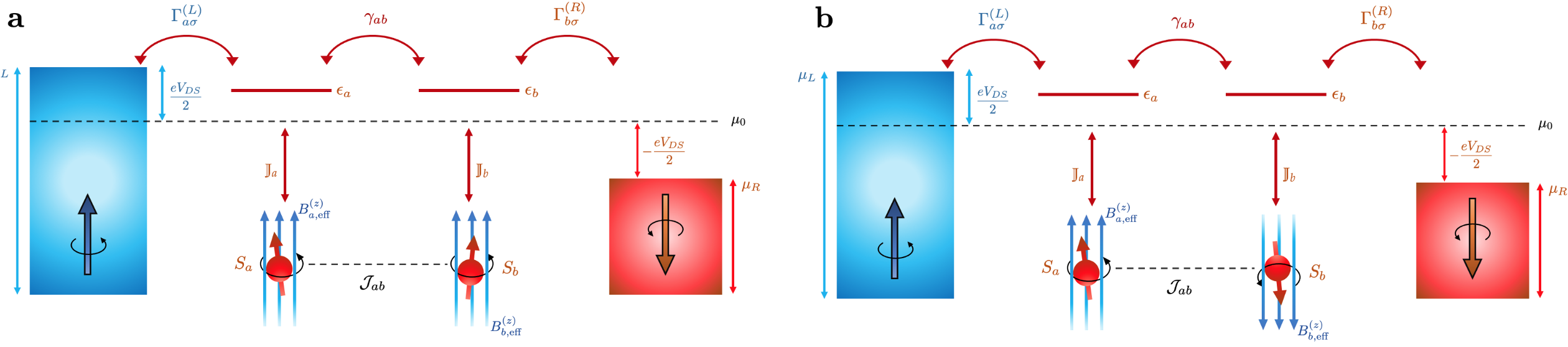}% Here is how to import EPS art
\caption{\label{dimer} Paramagnetic Dimer Embedded in a Ferromagnetic Tunnel Junction: A magnetic molecule consisting of two molecules with spin units $\bm{S}_{a}$ and $\bm{S}_{b}$ Kondo coupled to electrons in with energies $\epsilon_{a\sigma}$ and $\epsilon_{b\sigma}$ through interaction constants $\mathbb{J}_{a}$ and $\mathbb{J}_{b}$ respectively. $\gamma_{ab}$ is the hopping parameter between the energy levels and parameters $\Gamma^{(L)}_{a\sigma}$ and $\Gamma^{(R)}_{b\sigma}$ denote the coupling of levels $\epsilon_{a\sigma}$ and $\epsilon_{b\sigma}$ to the left and right ferromagnetic/metallic leads respectively. The junction in the figure has a bias protocol symmetric with respect to $\mu_{0}$ such that the chemical potentials in the left and right lead respectively depend on the bias voltage as $\mu_{l}=\mu_{0}+eV_{DS}/2$ and $\mu_{R}=\mu_{0}-eV_{DS}/2$.}
\end{figure*}

\begin{widetext}
\begin{align}
	\mathcal{\bm{H}}&=\mathcal{\bm{H}}_{\alpha}+\mathcal{\bm{H}}_{\beta}+\mathcal{\bm{H}}_{T}+\mathcal{\bm{H}}_{mol},
	\label{ham1}
	\\
	\mathcal{\bm{H}}_{\alpha}&=\sum_{\bm{k}\sigma}\epsilon_{\bm{k}\sigma}\bm{c}^{\dagger}_{\bm{k}\sigma}
	\bm{c}_{\bm{k}\sigma},
	~~\mathcal{\bm{H}}_{\beta}=\sum_{\bm{q}\sigma}\epsilon_{\bm{q}\sigma}\bm{c}^{\dagger}_{\bm{q}\sigma}
	\bm{c}_{\bm{q}\sigma},
	\label{ham3}
	\\
	\mathcal{\bm{H}}_{T}&=\sum_{\bm{k}\sigma}v^{(a)}_{\bm{k}\sigma}\bm{c}^{\dagger}_{\bm{k}\sigma}\bm{d}_{a\sigma}
	+v^{(a)*}_{\bm{k}\sigma}\bm{d}^{\dagger}_{a\sigma}\bm{c}_{\bm{k}\sigma} +\sum_{\bm{q}\sigma}v^{(b)}_{\bm{q}\sigma}\bm{c}^{\dagger}_{\bm{q}\sigma}\bm{d}_{b\sigma}
	+v^{(b)*}_{\bm{q}\sigma}\bm{d}^{\dagger}_{b\sigma}\bm{c}_{\bm{q}\sigma},
	\label{ham4}
	\\
    \mathcal{\bm{H}}_{mol}&=\sum_{\sigma}\epsilon_{a\sigma}\bm{d}^{\dagger}_{a\sigma}\bm{d}_{a\sigma}
	+\sum_{\sigma}\epsilon_{b\sigma}\bm{d}^{\dagger}_{b\sigma}\bm{d}_{b\sigma} +\mathbb{J}_{a}\bm{S}_{a}\cdot\sum_{\sigma\sigma'}\bm{d}^{\dagger}_{a\sigma}\bm{\sigma}_{\sigma\sigma'}\bm{d}_{a\sigma'}+\mathbb{J}_{b}\bm{S}_{b}\cdot\sum_{\sigma\sigma'}\bm{d}^{\dagger}_{b\sigma}\bm{\sigma}_{\sigma\sigma'}\bm{d}_{b\sigma'}+\gamma_{12}\sum_{\sigma}\left(\bm{d}^{\dagger}_{a\sigma}\bm{d}_{b\sigma}+\bm{d}^{\dagger}_{b\sigma}\bm{d}_{a\sigma}\right),
	\label{ham5}
\end{align}
\end{widetext}

where $\mathcal{\bm{H}}_{\alpha}$ and $\mathcal{\bm{H}}_{\beta}$ are the Hamiltonians corresponding to the ferromagnetic/metallic leads labeled with $\alpha$ and $\beta$, $\mathcal{\bm{H}}_{mol}$ is the Hamiltonian describing the energetics of the magnetic molecule which includes hoping and Kondo interaction terms and $\mathcal{\bm{H}}_{T}$ is the tunneling Hamiltonian, which encodes the energetics of electrons in the left lead tunneling to and from site $a$ of the molecule or of electrons tunneling to and from site $b$ of the same molecule. For the left ferromagnetic/metallic lead the wave vector of the electrons is labeled as $\bm{k}$, the spin is labeled as $\sigma$, and operator $\bm{c}^{\dagger}_{\bm{k}\sigma}$ $(\bm{c}_{\bm{k}\sigma})$ creates (annihilates) a single particle state at the energy band labeled as $\epsilon_{\bm{k}\sigma}$. For the right ferromagnetic/metallic lead, the wave vector is labeled as $\bm{q}$ and the spin is labeled as $\sigma$ as there are no spin flip processes, whereas all the other variables and parameters $\bm{c}^{\dagger}_{\bm{q}\sigma},\bm{c}_{\bm{q}\sigma},\epsilon_{\bm{q}\sigma}$ are as defined for the left lead. Operators $\bm{d}^{\dagger}_{a\sigma}$ $(\bm{d}_{a\sigma})$ and $\bm{d}^{\dagger}_{b\sigma}$ $(\bm{d}_{b\sigma})$ create (annihilate) a single particle in the energy level $\epsilon_{a\sigma}$ and $\epsilon_{b\sigma}$ respectively, which hybridize with an interaction strength $\gamma_{ab}$ hence allowing for particle hopping. The Kondo interaction strength appearing in the molecular Hamiltonian is labeled as $\mathbb{J}_{a}$ and $\mathbb{J}_{b}$ respectively for site $a$ and $b$, which couples the localized spin moment $\bm{S}_{a}$ ($\bm{S}_{b}$) with the spin of conduction electrons in energy level $\epsilon_{a\sigma}$ ($\epsilon_{b\sigma}$) denoted by the spin operator given by $\bm{s}^{(e)}_{a}=\sum_{\sigma\sigma'}\bm{d}^{\dagger}_{a\sigma}\bm{\sigma}_{\sigma\sigma'}\bm{d}^{\dagger}_{a\sigma}$ ($\bm{s}^{(e)}_{b}=\sum_{\sigma\sigma'}\bm{d}^{\dagger}_{b\sigma}\bm{\sigma}_{\sigma\sigma'}\bm{d}^{\dagger}_{b\sigma}$).
\\
Following references \cite{Fransson2010dyn,Fransson2014,Fransson2017}, an effective first order and seconder action can be defined for spin moments driven out of equilibrium which are given by:
\begin{align}
	\delta S^{(1)}&=\int dt \sum_{m}\bm{B}_{m,eff}(t)\cdot\bm{S}_{m}(t),
	\label{eff1}
	\\
	\delta S^{(2)}&=\int dtdt' \sum_{mn} \mathcal{J}^{R}_{mn}(t,t')\bm{S}_{m}(t)\cdot\bm{S}_{n}(t') \nonumber
	\\
        +\bm{T}^{R}_{mn} & (t,t')\cdot\bm{S}_{m}(t)\times\bm{S}_{n}(t')
	+\bm{S}_{m}(t)\cdot\mathbb{I}^{R}_{mn}(t,t')\cdot\bm{S}_{n}(t'),
	\label{eff2}
\end{align}
where $\bm{B}_{m,eff}(t)$ is the effective magnetic field acting on the spin moment $\bm{S}_{m}(t)$ which is created by the spin polarization of the surrounding electronic background, $\mathcal{J}^{R}_{mn}(t,t')$, $\bm{T}^{R}_{mn}(t,t')$ and $\mathbb{I}^{R}_{mn}(t,t')$ are the isotropic, anti-symmetric anisotropic and symmetric anisotropic components of the retarded effective magnetic susceptibility \cite{Fransson2010b,Secchi2013,Fransson2017} respectively in the rotated Keldysh space \cite{Keldysh1965}, which resemble the microscopic theory of spin magnetism derived by Toru Moriya \cite{Moriya1960}. These components of the effective spin action to first and second order are given in terms of the nonequilibrium charge and spin Green's functions $\mathcal{G}^{</>}(t,t')$ and $\bm{G}^{</>}(t,t')$ (note that $\bm{G}$ has components $G_{x}$, $G_{y}$ and $G_{z}$.) as specified in reference \cite{Hammar2016}, where in turn, in coincidence with reference \cite{Fransson2014}, shows that in the time invariant case:
\\
\\
\textsl{Symmetric Isotropic Effective Exchange Interaction}:

\begin{align}
	\mathcal{J}^{R}_{mn}=&\frac{\mathbb{J}_{a}\mathbb{J}_{b}}{2}\int\int \Big(\frac{\mathcal{G}_{mn}^{>}(\epsilon)\mathcal{G}_{nm}^{<}(\epsilon')-\mathcal{G}_{mn}^{<}(\epsilon)\mathcal{G}_{nm}^{>}(\epsilon')}{\epsilon-\epsilon'}\nonumber
	\\ 
	~~~~~~+&\frac{\bm{G}_{mn}^{>}(\epsilon)\cdot\bm{G}_{nm}^{<}(\epsilon')-\bm{G}_{mn}^{<}(\epsilon)\cdot\bm{G}_{nm}^{>}(\epsilon')}{\epsilon-\epsilon'}\Big) \frac{d\epsilon}{2\pi}\frac{d\epsilon'}{2\pi}
	\label{exch1}
\end{align}
\textsl{Anti-Symmetric Anisotropic (Chiral) Effective Exchange Interaction}:
\begin{align}
	\bm{T}^{R}_{mn}= & \frac{\mathbb{J}_{a}\mathbb{J}_{b}}{4} \int \Big( \mathcal{G}_{mn}^{<}(\epsilon)\bm{G}_{nm}^{>}(\epsilon)-\mathcal{G}_{mn}^{>}(\epsilon)\bm{G}_{nm}^{<}(\epsilon) \nonumber
    \\
    -& \bm{G}_{mn}^{<}(\epsilon)\mathcal{G}_{nm}^{>}(\epsilon)+\bm{G}_{mn}^{>}(\epsilon)\mathcal{G}_{nm}^{<}(\epsilon)\Big) \frac{d\epsilon}{2\pi}
	\label{exch2}
\end{align}
\textsl{Symmetric Anisotropic Effective Exchange Interaction}:
\begin{align}
	\mathbb{I}^{R}_{mn}=&\frac{\mathbb{J}_{a}\mathbb{J}_{b}}{2}\int\int \Big(\frac{\bm{G}_{mn}^{>}(\epsilon)\bm{G}_{nm}^{<}(\epsilon')-\bm{G}_{mn}^{<}(\epsilon)\bm{G}_{nm}^{>}(\epsilon')}{\epsilon-\epsilon'}\nonumber
     \\
    ~~~~~+&\frac{\bm{G}_{nm}^{>}(\epsilon')\bm{G}_{mn}^{<}(\epsilon)-\bm{G}_{nm}^{<}(\epsilon')\bm{G}_{nm}^{>}(\epsilon)}{\epsilon-\epsilon'}\Big)\frac{d\epsilon}{2\pi}\frac{d\epsilon'}{2\pi}
	\label{exch3}
\end{align}
where the Green's functions used in the evaluation of expressions \ref{exch1}, \ref{exch2} and \ref{exch3} have been calculated using the Keldysh equation \cite{Jauho1994,Haug2013} given by:
\begin{align}
	G^{</>}_{mn}(\omega)=\sum_{ij}G^{R}_{mi}(\omega)\Sigma^{</>}_{ij}(\omega)G^{A}_{jn}(\omega),
	\label{keldysh_eq}
\end{align}
from the knowledge of $G^{R}(\omega)$ which can be calculated from the equation of motion method as detailed in chapter 8 of reference \cite{VasquezJaramillo2018b}, and from the knowledge of the self-energy $\Sigma^{</>}(\omega)$ which accounts for the coupling of the electron sea hosting the spins with the external field driving the nonequilibrium interactions, and is often for the case of electronic reservoirs given in the wide-band limit \cite{Jauho1994} by:
\begin{align}
	\Sigma^{</>}(\omega)=(\pm i)\sum_{\kappa=L,R}f_{\kappa}(\pm\omega)\Gamma^{(\kappa)},
	\label{self1}
\end{align}
where $f_{\kappa}(\omega)$ is the occupation function of the reservoir labeled as $\kappa$ which drives the system of interest. For the specific case of the Hamiltonian given by expressions \ref{ham1}, \ref{ham3}, \ref{ham4} and \ref{ham5}, $\Gamma^{(L)}$ and $\Gamma^{(R)}$ are given by:
\begin{align}
	\Gamma^{(L)}=\left[\begin{array}{cc}\Gamma_{a\sigma}^{(L)}&0
		\\
		0&0\end{array}\right];~~~~
	\Gamma^{(R)}=\left[\begin{array}{cc}0&0
		\\
		0&\Gamma_{b\sigma}^{(R)}\end{array}\right],
	\label{self2}
\end{align}
giving the following expression for $\Sigma^{</>}(\omega)$:
\begin{align}
	\Sigma^{</>}(\omega)=(\pm i)
	\left[\begin{array}{cc}f_{L}(\pm\omega)\Gamma_{a\sigma}^{(L)}&0
		\\
		0&f_{R}(\pm\omega)\Gamma_{b\sigma}^{(R)}\end{array}\right],
	\label{self3}
\end{align}
The quantities in expressions \ref{exch1}, \ref{exch2} and \ref{exch3} defined the parameters for an effective spin Hamiltonian from the second order effective spin action given by expression \ref{eff2}, leading to:
\begin{align}
	\mathcal{\bm{H}}_{spin}&=\sum_{mn} \mathcal{J}^{R}_{mn}\bm{S}_{m}\cdot\bm{S}_{n}
	+\bm{T}^{R}_{mn}\cdot\bm{S}_{m}\times\bm{S}_{n}
	+\bm{S}_{m}\cdot\mathbb{I}^{R}_{mn}\cdot\bm{S}_{n},
	\label{eff_ham1}
\end{align}
where for the case of a spin dimer rather reads:
\begin{align}
	\mathcal{\bm{H}}_{spin}&=\mathcal{J}^{R}_{ab}\bm{S}_{a}\cdot\bm{S}_{b}
	+\bm{T}^{R}_{ab}\cdot\bm{S}_{a}\times\bm{S}_{b}
	+\bm{S}_{a}\cdot\mathbb{I}^{R}_{ab}\cdot\bm{S}_{b}.
	\label{eff_ham2}
\end{align}
The effective Hamiltonian given by expression \ref{eff_ham2} yields zero expectation value for both spins $\bm{S}_{a}$ and $\bm{S}_{b}$, for vanishing spin orbit torques $\bm{T}^{R}_{ab}$ and Symmetric Anisotropic Exchange $\mathbb{I}^{R}_{ab}$.
%%%%%%%%%%%%%%%%%%%%%%%%%%%%%%%%%%%%%%%%%%%%%%%%%%%%%%%%%%%%%%%%%%%%%%%%%%%%%%%%%%%%%%%%%%%%%%%%%%%%%%%%%%%%%%%%%%%%%%%%%%%%%%%%%%%%%%%%%%%%%%%%%%%%%%%%%%%%%%%%%%%%%%%%%%%%%%%%%%%%%%%%%%%%%%%%%%%%%%%%%%%%%%%%%%%%%%%%%%%%%%%%%%%%%%%%%%%%%%%%%%%%%%%%%%%%%%%%%%%%%%%%%%%%%%%%%%%%%%%%%%%%%%%%%%%%%%%%%%%%%%%%%%%%%%%%%%%%%%%%%%%%%%%%%%%%%%%%%%%%%%%%%%%%%%%%%%%%%%%%%%%%%%%%%%%%%%%%%%%%%%%%%%%%%%%%%%%%%%%%
\section{Results and Discussion}
Here, we are interested in showing that, a symmetry breaking interaction in required to induced finite local magnetizations on single/individual spin units in magnetic molecules or quantum dots driven out of equilibrium. First, let's understand the voltage dependent eigenvalue problem presented here, and what is the voltage induced switching mechanisms in the isotropic and chiral exchange interactions as a function of magnetic driving, and then, let's argue why magnetic driving is necessary for local magnetization emergence.
\subsection{Voltage Dependent Parameters}
\begin{figure}[!ht]
	\centering
    \includegraphics[width=1.\columnwidth]{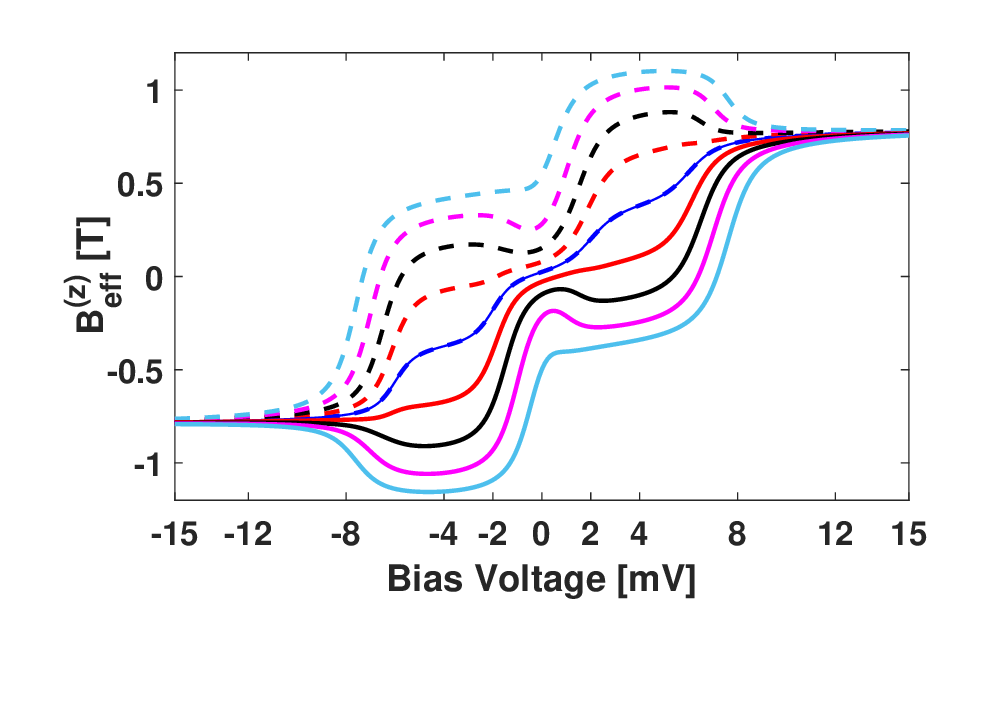}
	\caption{Effective Magnetic Field (Bird Shape): in the illustration we can appreciate the asymptotic behavior of the effective magnetic field induced in each units of the dimer. The dotted lines represents the magnetization of the left spin unit and the continuous lines represent the magnetization of the right spin unit. The dark blue represents the zero external magnetization condition. The red lines represent external magnetic fields of magnitud $1T$, the black the purple and the light blue represent correspondingly magnetic fields of magnitudes $3T$, $4T$ and $5T$.}
	\label{beff}
\end{figure}

\begin{figure}[!ht]
	\centering
	\includegraphics[width=1.\columnwidth]{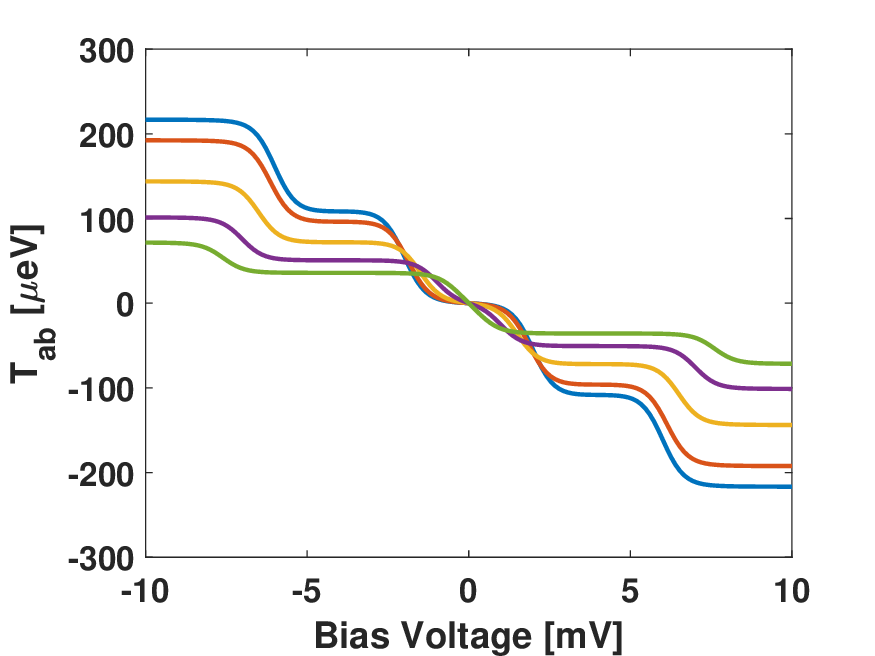}
	\caption{Chiral Anisotropic exchange as a function of external magnetizations for AMD protocol. The blue curve represents the zero external magnetization condition, while the red, yellow, purple and green represent the external magnetization strengths of $3T$, $4T$, $5T$ and $7T$ respectively.}
	\label{DM1}
\end{figure}

Departing from the Hamiltonian given by expression \ref{eff_ham2}, herein we consider two possible cases for its voltage dependent parameters given by expressions \ref{exch1} and \ref{exch2}, parallel magnetic driving (PMD) which is the case where both units of the spin dimer are magnetized in the same direction and with the same strength, and anti-parallel magnetic driving (AMD) where both spin units are magnetized with the same strength in opposite directions, that is, the magnetic fields magnetizing the dimer are staggered. Notice that for the case of a spin half dimer, the voltage dependent parameter given by expression \ref{exch3} does not contribute to the total energy of the spin Hamiltonian. Here, to break the spin state symmetry of the Hamiltonian, a magnetic field is required. In the absence of external magnetic fields, an effective magnetic field given by expression \ref{eff1} is generated when the spin asymmetry from the leads is transfer by a given mechanism into the dimer. For the case of staggered magnetic fields (AMD), the effective magnetic field shows an asymptotic behavior shown in fig. \ref{beff}:\\
where the dotted lines represents the effective magnetization of the level coupled to the left lead, and the continuous lines represent the effective magnetization of the level coupled to the right lead. Despite the asymptotic behavior of the magnetic field, the total magnetization is not zero as required for a paramagnetic dimer and as argued in \cite{Saygun2016a,Vasquez-Jaramillo2017}.
\begin{figure}[!ht]
	\centering
	\includegraphics[width=1.\columnwidth]{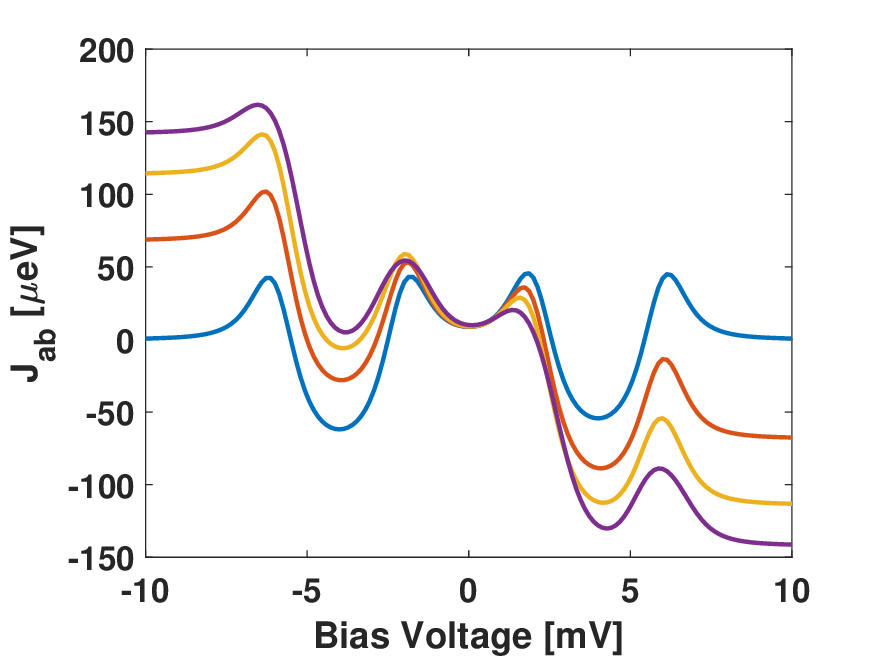}
	\caption{Isotropic exchange as a function of external magnetizations for PMD protocol. The blue curve represents the zero external magnetization condition, while the red, yellow and purple represent the external magnetization strengths of $3T$, $4T$ and $5T$ respectively.}
	\label{exchange}
\end{figure}
The parameters of the spin Hamiltonian can be analyzed for both driving protocols. For the case of parallel magnetic driving, the isotropic exchange as a function of different external magnetization conditions is shown in fig. \ref{exchange}.
\begin{figure}[!ht]
	\centering
	\includegraphics[width=1\columnwidth]{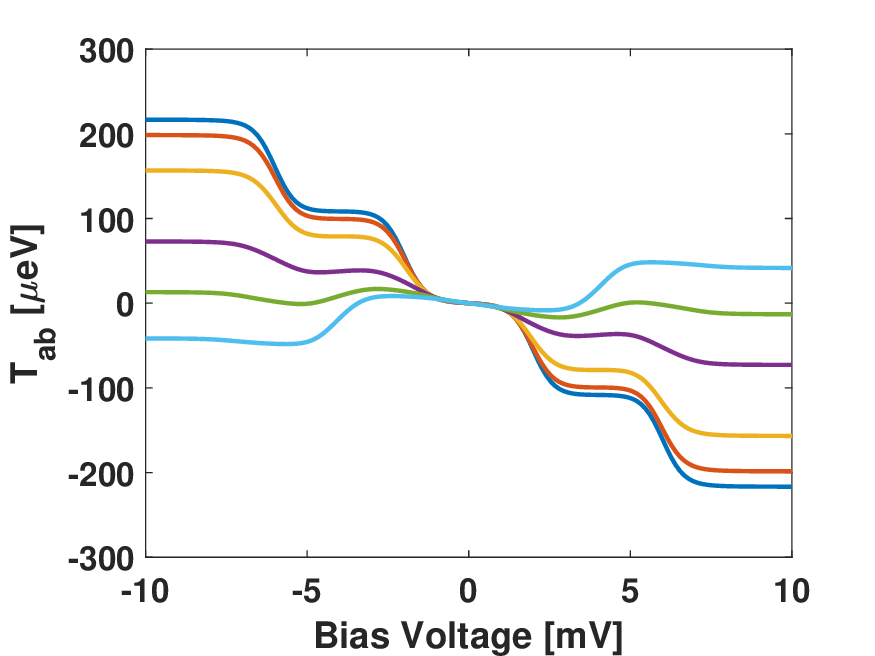}
	\caption{Chiral Anisotropic exchange as a function of external magnetizations for PMD protocol. The blue curve represents the zero external magnetization condition, while the red, yellow, purple, green and light blue represent the external magnetization strengths of $3T$, $4T$, $5T$, $7T$ and $8T$ respectively.}
	\label{DM}
\end{figure}
The chiral exchange interaction within the (PMD) protocol is shown in fig. \ref{DM}, Where we can appreciate the commutation between positive and negative chirality for large magnetic fields to finally vanish for even larger magnetic fields. Notice the difference in units of the exchange interactions presented here ($\mu eV$) with the ones presented in \cite{Fransson2014}, a fact that is basically due to the proper normalization of the nonequilibrium Green's function in the numerical evaluation of the exchanges.
\begin{figure}[!ht]
	\centering
	\includegraphics[width=1.\columnwidth]{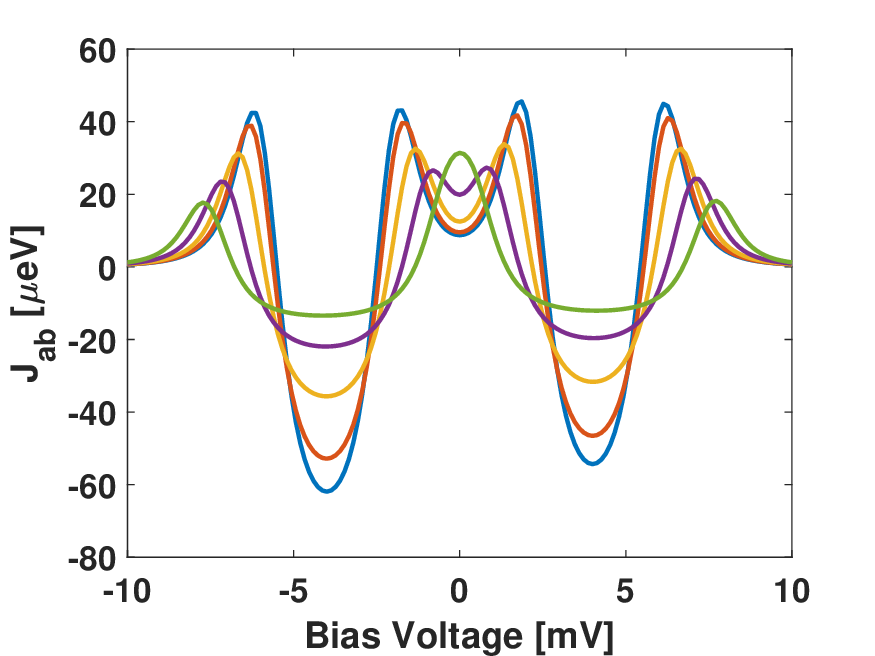}
	\caption{Isotropic exchange as a function of external magnetizations for AMD protocol. The blue curve represents the zero external magnetization condition, while the red, yellow, purple and green represent the external magnetization strengths of $3T$, $4T$, $5T$ and $7T$ respectively.}
	\label{exchange1}
\end{figure}
For the case of AMD, the isotropic and chiral anisotropic exchange are respectively shown in figs \ref{exchange1} and \ref{DM1}.
A fundamental difference between both driving protocols is that, at large voltage bias the PMD protocol produces constant exchange, and locks the magnetic state of the molecule to a singlet configuration, while the AMD protocol produces zero exchange at large bias yielding a four fold degeneracy and a notable Kondo like peak for large enough magnetic fields. In the chiral anisotropic exchange interaction one can see that under AMD protocol, the chirality of the molecule is more stable as a function of bias voltage as compare with the PMD protocol. At the end the choice of protocol will depend more on the asymptotic behavior of the effective magnetic field which will produce a zero net magnetization in the molecule, for such reason we will stick to AMD protocol.
\\
\subsection{Local Magnetizations of Individual Spin Units}
We consider the thermal average of individual spins in the dimer. The thermal expectation value of an arbitrary spin operator in an arbitrary direction is given by:
\begin{equation}
	\left\langle \bm{S}^{(\lambda)}_{j}\right\rangle=\frac{\texttt{Tr}\left(e^{-\beta\mathcal{\bm{H}}_{spin}}\bm{S}_{j}\right)}{\mathcal{Z}}.
	\label{expec1}
\end{equation}
To evaluate expression \ref{expec1}, one must define the eigen-basis $\ket{\phi_{n}}$ and the corresponding eigen-energies. In the present case, the eigen-basis are chosen as the eigenstates of the Hamiltonian $\mathcal{\bm{H}}_{spin}$, that is, $\ket{0}=\chi_{s}$, $\ket{1}=\chi^{(1)}_{t}$, $\ket{2}=\chi^{(2)}_{t}$ and 
$\ket{3}=\chi^{(3)}_{t}$, and with corresponding eigen-energies given by: $\mathcal{E}_{0}=-3\mathcal{J}_{AB}$ and $\mathcal{E}_{1,2,3}=\mathcal{J}_{AB}$. Now, the partition function $\mathcal{Z}$ can be evaluated as follows:

\begin{align}
	\mathcal{Z}&=\texttt{Tr}\left(e^{-\beta\mathcal{\bm{H}}_{spin}}\right)=\sum_{n}\braket{n|e^{-\beta\mathcal{\bm{H}}_{spin}}|n}
	=\sum_{n}e^{-\beta\mathcal{E}_{n}},
	\nonumber
	\\
	&=e^{-\beta\mathcal{E}_{0}}+e^{-\beta\mathcal{E}_{1}}+e^{-\beta\mathcal{E}_{2}}
	+e^{-\beta\mathcal{E}_{3}}=e^{3\beta\mathcal{J}_{AB}}+3e^{-\beta\mathcal{J}_{AB}}.
	\label{expec2}
\end{align}
Moreover, to evaluate $\left\langle \bm{S}^{(z)}_{a}\right\rangle$ and $\left\langle\bm{S}^{(z)}_{b}\right\rangle$, we evaluate the following traces: $\texttt{Tr}\left(e^{-\beta\mathcal{\bm{H}}_{spin}}\bm{S}^{(z)}_{a}\right)$ and $\texttt{Tr}\left(e^{-\beta\mathcal{\bm{H}}_{spin}}\bm{S}^{(z)}_{b}\right)$. First let's evaluate the thermal trace with respect to $\bm{S}^{(z)}_{a}$:
\begin{align} \nonumber
	\texttt{Tr}\left(e^{-\beta\mathcal{\bm{H}}_{spin}}\bm{S}^{(z)}_{a}\right)&=\sum_{n}\braket{n|e^{-\beta\mathcal{\bm{H}}_{spin}}\bm{S}^{(z)}_{a}|n}
 \\
 &=\sum_{n}e^{-\beta\mathcal{E}_{n}}\braket{n|\bm{S}^{(z)}_{a}|n},
	\nonumber
	\\
	&=e^{-\beta\mathcal{E}_{0}}\braket{0|\bm{S}^{(z)}_{a}|0}+e^{-\beta\mathcal{E}_{1}}\braket{1|\bm{S}^{(z)}_{a}|1} \nonumber
	\\
 &~~~~~+e^{-\beta\mathcal{E}_{2}}\braket{2|\bm{S}^{(z)}_{a}|2}+e^{-\beta\mathcal{E}_{3}}\braket{3|\bm{S}^{(z)}_{a}|3}.
	\label{expec3}
\end{align}

The terms $\langle 0|\bm{S}^{(z)}_{a}|0 \rangle$, $\langle 1|\bm{S}^{(z)}_{a}|1\rangle$, $\langle 2|\bm{S}^{(z)}_{a}|2\rangle$ and $\langle 3|\bm{S}^{(z)}_{a}|3 \rangle$ in expression \ref{expec3} can be evaluated in the following form:
where the properties $\bm{S}^{(z)}_{a}|\uparrow(\uparrow,\downarrow)\rangle=\hbar/2~ |\uparrow(\uparrow,\downarrow)\rangle$ and $\bm{S}^{(z)}_{a}|\downarrow(\uparrow,\downarrow)\rangle=-\hbar/2 ~ |\downarrow(\uparrow,\downarrow)\rangle$ have been used. Replacing expressions \ref{expec4}, \ref{expec5}, \ref{expec6} and \ref{expec7} from the supporting information in expression \ref{expec3}, the above mentioned trace reads:
\begin{align}
	\texttt{Tr}\left(e^{-\beta\mathcal{\bm{H}}_{spin}}\bm{S}^{(z)}_{a}\right)&=e^{-\beta\mathcal{J}_{ab}}\left(\frac{\hbar}{2}\right)+e^{-\beta\mathcal{J}_{ab}}\left(-\frac{\hbar}{2}\right)=0,
	\label{expec8}
\end{align}
therefore the expectation value $\left\langle \bm{S}^{(z)}_{a}\right\rangle$ vanishes:
\begin{align}
	\left\langle \bm{S}^{(z)}_{a}\right\rangle=\frac{\hbar}{2}\frac{e^{-\beta\mathcal{J}_{ab}}-e^{-\beta\mathcal{J}_{ab}}}{e^{3\beta\mathcal{J}_{ab}}+3e^{-\beta\mathcal{J}_{ab}}}=0.
	\label{expec9}
\end{align}
To induce finite values for $\left\langle \bm{S}^{(z)}_{a}\right\rangle$, a symmetry must be broken such that the eigen-energies corresponding to the eigenstates $\ket{2}$ and $\ket{3}$ should be different, that is, $\mathcal{E}_{2}\neq\mathcal{E}_{3}$, which gives the following value for $\left\langle \bm{S}^{(z)}_{a}\right\rangle$:
\begin{align}
	\left\langle \bm{S}^{(z)}_{a}\right\rangle=\frac{\hbar}{2}\frac{e^{-\beta\mathcal{E}_{2}}-e^{-\beta\mathcal{E}_{3}}}{e^{3\beta\mathcal{J}_{ab}}+3e^{-\beta\mathcal{J}_{ab}}}.
	\label{expec10}
\end{align}
For the case of the expectation value of the second molecular spin, $\left\langle \bm{S}^{(z)}_{b}\right\rangle$, it can be straightforward seen that expressions \ref{expec4}, \ref{expec5}, \ref{expec6} and \ref{expec7} also apply to this case, hence giving:
\begin{align}
	\left\langle \bm{S}^{(z)}_{b}\right\rangle=\frac{\hbar}{2}\frac{e^{-\beta\mathcal{J}_{ab}}-e^{-\beta\mathcal{J}_{ab}}}{e^{3\beta\mathcal{J}_{ab}}+3e^{-\beta\mathcal{J}_{ab}}}=0,
	\label{expec11}
\end{align}
and in the presence of axial symmetry breaking (magnetic fields), it rather gives:
\begin{align}
	\left\langle \bm{S}^{(z)}_{b}\right\rangle=\frac{\hbar}{2}\frac{e^{-\beta\mathcal{E}_{2}}-e^{-\beta\mathcal{E}_{3}}}{e^{3\beta\mathcal{J}_{ab}}+3e^{-\beta\mathcal{J}_{ab}}}.
	\label{expec12}
\end{align}
for the particular case of the PMD protocol, in which the eigenvalues $\mathcal{E}_{2}$ and $\mathcal{E}_{3}$ are given by expression \ref{expec22}, the expectation values $\left\langle \bm{S}^{(z)}_{a}\right\rangle$ and $\left\langle \bm{S}^{(z)}_{b}\right\rangle$:
\begin{align}
	\hspace{-1.0cm}\langle{{\bf{S}}_{a}}^{(z)}\rangle=\frac{\hbar\sinh\left[\frac{ \Delta B_{z}}{2k_{B}T}\right]}{\left(1+2\cosh\left[\frac{\Delta B_{z}}{2k_{B}T}\right]+e^{\beta\mathcal{J}_{ab}}\right)}, \nonumber
 \\
 \langle{{\bf{S}}_{b}}^{(z)}\rangle=\frac{\hbar\sinh\left[\frac{\Delta B_{z}}{2k_{B}T}\right]}{\left(1+2\cosh\left[\frac{\Delta B_{z}}{2k_{B}T}\right]+e^{\beta\mathcal{J}_{ab}}\right)},
	\label{expec12a}
	\end{align}

\begin{figure*}
\includegraphics[width=1\textwidth]{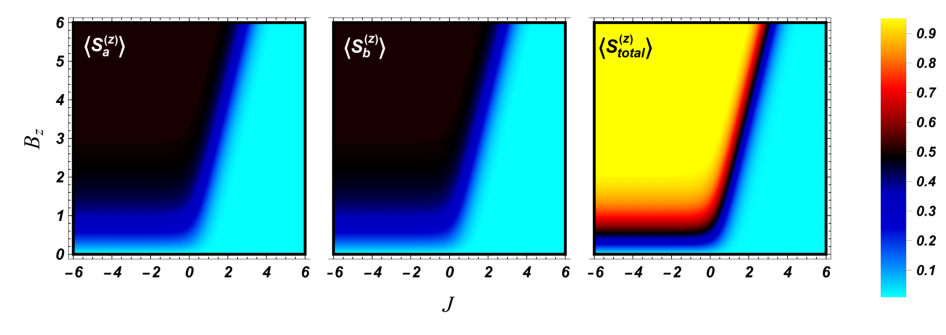}% Here is how to import EPS art
	\caption{Magnetic moment per spin unit in a molecular dimer and the total spin moment of the molecule under Parallel Magnetic Driving (PMD) Protocol at $T=4K$.}
	\label{moment1}
\end{figure*}

\begin{figure*}
\includegraphics[width=1\textwidth]{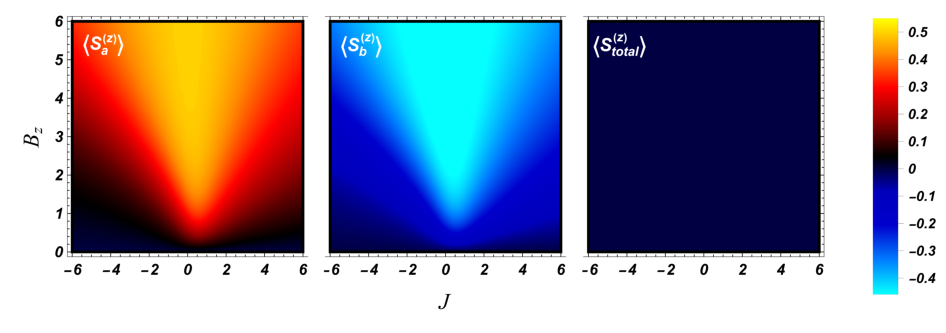}% Here is how to import EPS art
	\caption{Magnetic moment per spin unit in a molecular dimer and the total spinmoment of the molecule under Anti-Parallel Magnetic Driving (AMD) Protocol at $T=4K$.}
	\label{moment2}
\end{figure*}
where it can be notice that, for finite spin expectation values it is required to have finite magnetic induction fields and relatively low temperature compare to the Zeeman energy (See that by Taylor expanding the hyperbolic functions, Curie's law is retrieved). For details see supporting information.	

%%%%%%%%%%ANEXO%%%%%%%%%%%%%%%%%%%%%%%%%%%

\subsection{Breaking Time Reversal Symmetry: The Zeeman Energy}
For the case in which magnetic fields are present in the geographical vecinity of the molecule, the spin Hamiltonian reads:

\begin{align}
	\mathcal{\bm{H}}_{spin}=\mathcal{\bm{H}}_{Heiss}+\mathcal{\bm{H}}_{Zeeman},
	\label{expec13}
\end{align}

where $\mathcal{\bm{H}}_{Heiss}$ is the Hamiltonian we considered in the previous section, which is $SO_{3}$ invariant hence producing zero net magnetic moment, and $\mathcal{\bm{H}}_{Zeeman}$ is the Hamiltonian corresponding to the Zeeman energy of each of the spins represented, and the latter and the former are given by:
\begin{align}
	\mathcal{\bm{H}}_{Heiss}&=\mathcal{J}_{ab}\bm{S}_{a}\cdot\bm{S}_{b},
	\\
	\mathcal{\bm{H}}_{Zeeman}&=-g_{a}\mu_{B}\bm{B}_{a,eff}\cdot\bm{S}_{a}-g_{b}\mu_{B}\bm{B}_{b,eff}\cdot\bm{S}_{b},
	\label{expec14}
\end{align}
where the constants $g_{i}$ and $\mu_{B}$ are respectively the giromagnetic ratio for $i=a,b$ and the Bohr's magneton. $\bm{B}_{i,eff}$ is the effective magnetic field acting on spin $\bm{S}_{i}$ for $i=a,b$, which is a combination of the externally applied magnetic field and the magnetic field created by the spin polarization surrounding molecule $i$. For the specific case in which the magnetic fields on molecules $a$ and $b$ are pointing along the $z-$ direction, the Zeeman energy is the represented by the following Hamiltonian:
\begin{align}
	\mathcal{\bm{H}}_{Zeeman}&=-g_{a}\mu_{B}B^{(z)}_{a,eff}\bm{S}^{(z)}_{a}-g_{b}\mu_{B}B^{(z)}_{b,eff}\bm{S}^{(z)}_{b}.
	\label{expec15}
\end{align}
In matrix form, $\bm{S}^{(z)}_{a}=S^{(z)}_{a}\otimes \mathbb{I}_{2\times 2}$ and $\bm{S}^{(z)}_{b}=\mathbb{I}_{2\times 2}\otimes S^{(z)}_{b}$, which explicitly reads:

\begin{align}
	\bm{S}^{(z)}_{a}=\left[\begin{array}{cccc}1&0&0&0
		\\
		0&1&0&0
		\\
		0&0&-1&0
		\\
		0&0&0&-1
	\end{array}\right];~~~~~
	\bm{S}^{(z)}_{b}=\left[\begin{array}{cccc}1&0&0&0
		\\
		0&-1&0&0
		\\
		0&0&1&0
		\\
		0&0&0&-1
	\end{array}\right];
\end{align}

from where the Zeeman energy term given by expression \ref{expec15} is of diagonal form, and its matrix elements are given by:
\begin{align}
\mathcal{\bm{H}}_{Zeeman}^{(11)}&=-g_{a}\mu_{B}B^{(z)}_{a,eff}-g_{b}\mu_{B}B^{(z)}_{b,eff}
\nonumber
\\
\mathcal{\bm{H}}_{Zeeman}^{(22)}&=-g_{a}\mu_{B}B^{(z)}_{a,eff}+g_{b}\mu_{B}B^{(z)}_{b,eff}  
\nonumber
\\
\mathcal{\bm{H}}_{Zeeman}^{(33)}&=g_{a}\mu_{B}B^{(z)}_{a,eff}-g_{b}\mu_{B}B^{(z)}_{b,eff}
\nonumber
\\
\mathcal{\bm{H}}_{Zeeman}^{(44)}&=g_{a}\mu_{B}B^{(z)}_{a,eff}+g_{b}\mu_{B}B^{(z)}_{b,eff}~~  
\label{expec16}
\end{align}
In this paper we consider two magnetization protocols in the paramagnetic dimer. One is a protocol in which both molecules are externally magnetized in the same direction, with the same strength, and we will refer to this mode of operation as parallel magnetic driving (PMD, $B^{(z)}_{a,eff}=B^{(z)}_{b,eff}$). The second protocol namely anti-parallel magnetic driving (AMD, $B^{(z)}_{a,eff}=-B^{(z)}_{b,eff}$), consists in magnetizing both molecules in opposite directions with equal strengths. To generalize both protocols, we impose the relation $B^{(z)}_{b,eff}=\xi B^{(z)}_{a,eff}=\xi B^{(z)}_{b,eff}$, where $\xi=1$ for PMD and $\xi=-1$ for AMD, from where, the Zeeman Hamiltonian can be written more compactly by defining $\gamma_{\pm}(\xi)=g_{a}\pm\xi g_{b}$:

\begin{widetext}
\begin{align}
	\mathcal{\bm{H}}_{Zeeman}(\xi)&=\left[\begin{array}{cccc}-\gamma_{+}(\xi)\mu_{B}B^{(z)}_{eff}&0&0&0
		\\
		0&-\gamma_{-}(\xi)\mu_{B}B^{(z)}_{eff}&0&0
		\\
		0&0&\gamma_{-}(\xi)\mu_{B}B^{(z)}_{eff}&0
		\\
		0&0&0&\gamma_{+}(\xi)\mu_{B}B^{(z)}_{eff}
	\end{array}\right].
	\label{expec17}
\end{align}
\end{widetext}
Finally, the spin Hamiltonian can be written in matrix form by replacing expression \ref{expec17} into expression \ref{expec13}, giving:
\begin{widetext}
\begin{align}
	\mathcal{\bm{H}}_{spin}(\xi)&=\left[\begin{array}{cccc}\mathcal{J}_{ab}-\gamma_{+}(\xi)\mu_{B}B^{(z)}_{eff}&0&0&0
		\\
		0&-\mathcal{J}_{ab}-\gamma_{-}(\xi)\mu_{B}B^{(z)}_{eff}&2\mathcal{J}_{ab}&0
		\\
		0&2\mathcal{J}_{ab}&-\mathcal{J}_{ab}+\gamma_{-}(\xi)\mu_{B}B^{(z)}_{eff}&0
		\\
		0&0&0&\mathcal{J}_{ab}+\gamma_{+}(\xi)\mu_{B}B^{(z)}_{eff}
	\end{array}\right].
	\label{expec18}
\end{align}
\end{widetext}
The two cases of interest are $\xi=1$ for PMD and $\xi=-1$ for AMD, which give the following two spin Hamiltonians:

\begin{widetext}
\begin{align}
	\mathcal{\bm{H}}_{spin}(\xi=1)&=\left[\begin{array}{cccc}\mathcal{J}_{ab}-2g\mu_{B}B^{(z)}_{eff}&0&0&0
		\\
		0&-\mathcal{J}_{ab}&2\mathcal{J}_{ab}&0
		\\
		0&2\mathcal{J}_{ab}&-\mathcal{J}_{ab}&0
		\\
		0&0&0&\mathcal{J}_{ab}+2g\mu_{B}B^{(z)}_{eff}
	\end{array}\right],
	\label{expec20}
\end{align}
\begin{align}
	\mathcal{\bm{H}}_{spin}(\xi=-1)&=\left[\begin{array}{cccc}\mathcal{J}_{ab}&0&0&0
		\\
		0&-\mathcal{J}_{ab}-2g\mu_{B}B^{(z)}_{eff}&2\mathcal{J}_{ab}&0
		\\
		0&2\mathcal{J}_{ab}&-\mathcal{J}_{ab}+2g\mu_{B}B^{(z)}_{eff}&0
		\\
		0&0&0&\mathcal{J}_{ab}
	\end{array}\right].
	\label{expec21}
\end{align}
\end{widetext}
Moreover, we are interested in determining the spin excitation spectra for models given equations \ref{expec20} and \ref{expec21}. For expression \ref{expec20} (PMD protocol), the following eigenvalue $(\mathcal{E})$ problem is required to be solved (with $\Delta=2g\mu_{B}$, see details in supporting information):
\begin{align}
	|\mathcal{\bm{H}}_{spin}(\xi=1)-\mathcal{E}\mathbb{I}|&=0,
\end{align}
yielding the following eigenvalues:
\begin{align}
	\mathcal{E}_{0}&=-3\mathcal{J}_{ab};~~~~~~~~~~~~~\mathcal{E}_{1}=\mathcal{J}_{ab}; \nonumber
        \\
	\mathcal{E}_{2}&=\mathcal{J}_{ab}+\Delta B^{(z)}_{eff};~~~~\mathcal{E}_{3}=\mathcal{J}_{ab}-\Delta B^{(z)}_{eff};
	\label{expec22}
\end{align}

\begin{figure}[!ht]
	\centering
	\includegraphics[width=0.45\textwidth]{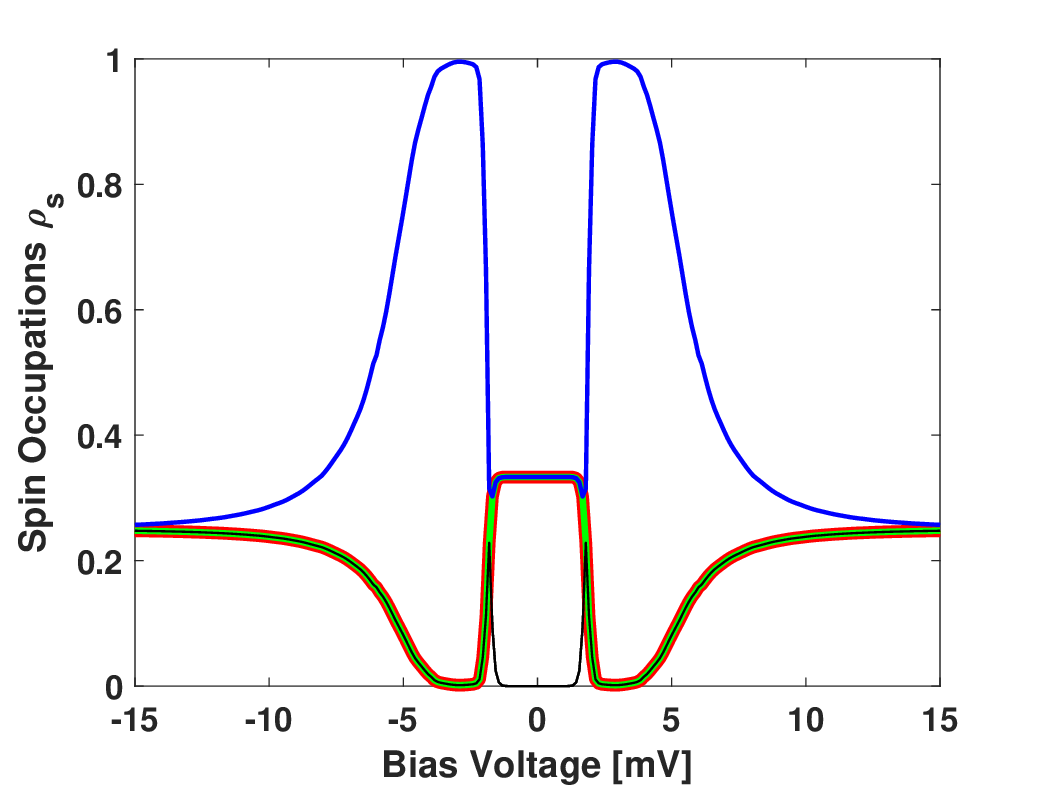}
	\caption{Spin excitation spectra in the absence of symmetry breaking: Here, the voltage dependent spin occupations from the Heisenberg Hamiltonian are plotted, where it can be clearly distinguished the triplet degenerate and the singlet regimes}
	\label{spin1}
\end{figure}

\begin{figure}[!ht]
	\centering
	\includegraphics[width=0.45\textwidth]{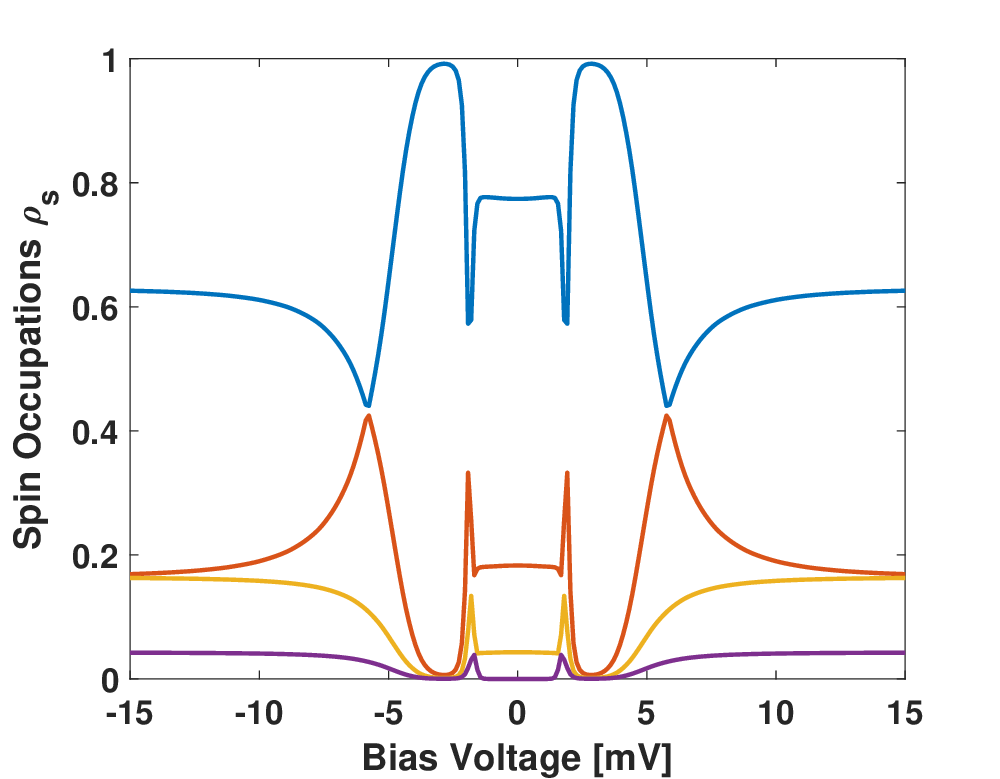}
	\caption{Spin Excitation Spectra for Parallel Magnetic Driving (PMD): Here, the voltage dependent spin occupations from the Heisenberg Hamiltonian are plotted in the presence of a symmetry breaking field, hence, showing the lifting of the triplet degeneracy, allowing for perfect distinction between the magnetic projections of each eigenstate of the spin Hamiltonian}
	\label{spin3}
\end{figure}

Now we proceed with the solution of the eigenvalue problem corresponding to expression \ref{expec21}:
\begin{align}
	|\mathcal{\bm{H}}_{spin}(\xi=-1)-\mathcal{E}\mathbb{I}|&=0,
	\label{expec23}
\end{align}
from where the eigenvalues corresponding to states $\ket{\uparrow\uparrow}$ and $\ket{\downarrow\downarrow}$ are given by:
\begin{align}
	\mathcal{E}_{0}=\mathcal{E}_{1}=\mathcal{J}_{ab},
	\label{expec24}
\end{align}
and the eigenvalues $\mathcal{E}_{2}$ and $\mathcal{E}_{3}$ read:
\begin{align}
	\mathcal{E}_{2,3}&=-\mathcal{J}_{ab}\pm\sqrt{4\mathcal{J}^{2}_{ab}+\left(\Delta B^{(z)}_{eff}\right)^{2}},
	\label{expec26}
\end{align}
corresponding to states:
$$
\ket{\chi_{s}}=\frac{1}{\sqrt{2}}\left(\ket{\uparrow\downarrow}-\ket{\downarrow\uparrow}\right)
$$
$$
\ket{\chi_{t}}=\frac{1}{\sqrt{2}}\left(\ket{\uparrow\downarrow}+\ket{\downarrow\uparrow}\right)
$$
$$
\ket{\chi_{t}}=\ket{\uparrow\uparrow}
$$
$$
\ket{\chi_{t}}=\ket{\downarrow\downarrow}
$$
\begin{figure}[!ht]
	\centering
	\includegraphics[width=0.45\textwidth]{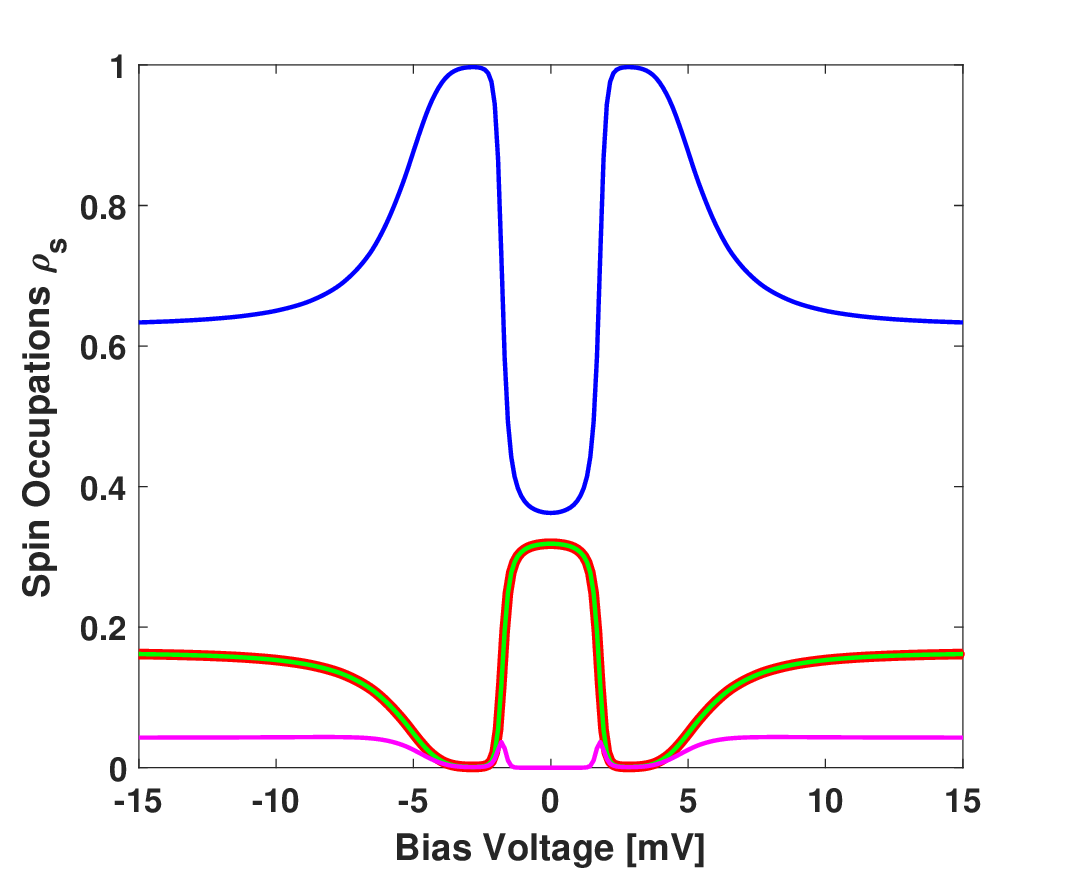}
	\caption{Spin Excitation Spectra for Anti-Parallel Magnetic Driving: Here, we plot the voltage dependent spin occupations of the Heisenberg Hamiltonian in the presence of staggered magnetic fields applied to each spin unit. It can be clearly seen that the entangled states are mainly non-degenerate}
	\label{spin4}
\end{figure}
for small effective magnetic fields $ B^{(z)}_{eff}$, expression \ref{expec26} can be approximated by a first order Taylor expansion as follows:
\begin{align}
	\mathcal{E}_{2}&\approx \mathcal{J}_{ab}+\frac{\left(\Delta B^{(z)}_{eff}\right)^{2}}{4\mathcal{J}_{ab}},
	\label{expec28}
	\\
	\mathcal{E}_{3}&\approx -3\mathcal{J}_{ab}-\frac{\left(\Delta B^{(z)}_{eff}\right)^{2}}{4\mathcal{J}_{ab}}.
	\label{expec29}
\end{align}

In relation to the voltage dependent spin excitation spectra, from figure \ref{spin1}, it can be clearly observed that an applied voltage does not lift the spin degeneracy, in the particular case where the electrodes are not ferromagnetic. Around zero bias, a spin triplet is formed as a consequence of the negativity of the nonequilibrium magnetic exchange, and the occupation for spin singlet is zero. In the case in which the absolute value of the bias voltage is far larger than zero, a spin singlet will be occupied with unit probability, and hence, the spin triplet will not be occupied at all. For very large absolute values of the bias voltage, where the nonequilibrium exchange vanishes, there will be a four fold spin degeneracy. To lift the spin degeneracy, we consider ferromagnetic electrodes and the PMD protocol, where the source of magnetization can be external or by proximity, and in figure \ref{spin3}, it is clearly seen that this symmetry breaking lifts the triplet degeneracy in the zero bias region and for larger absolute values of the voltages, even, lifting the four fold degeneracy. The spin degeneracy in the states with maximum and minimum spin projections remains in the presence of the application of the AMD protocol, though, the state for $s=1$ and $m_{s}=0$ has its spin degeneracy lifted under this protocol as it can be appreciated in figure \ref{spin4}.

\subsection{Numerical Evaluation of  Local Magnetizations and Quantum Transport}
\begin{figure}[!ht]
	\centering
	\includegraphics[width=0.45\textwidth]{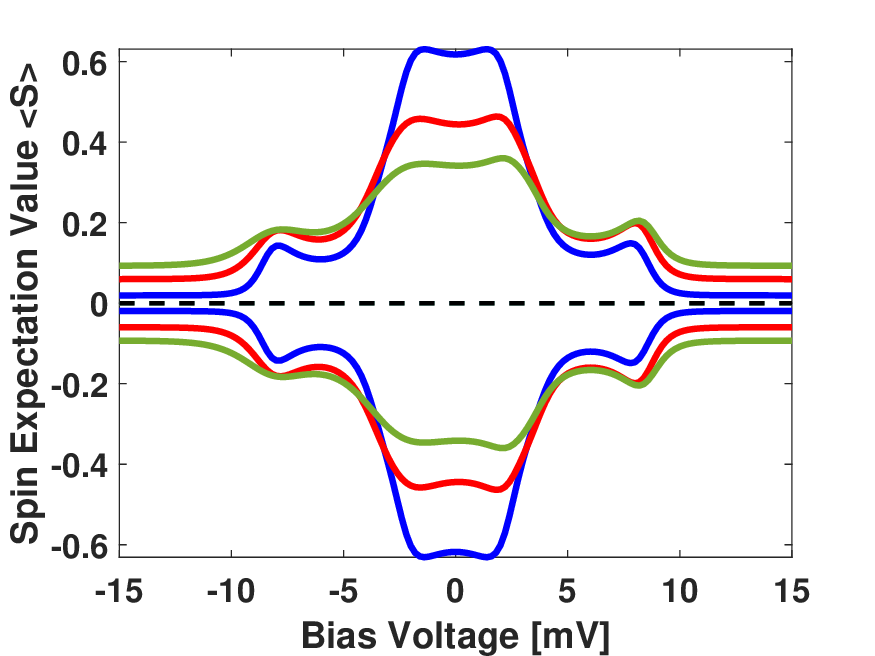}
	\caption{Illustration of the spin expectation value of each unit which diverges asymptotically from the zero (black dotted line). For the spin unit coupled to the left lead, the spin expectation value is positive as the local applied magnetic field is in the positive z-direction. For the spin unit coupled to the right lead, the spin expectation value is the negative because of the direction of the local applied magnetic field. As such, the AMD protocol produces equal but opposite in sign individual spin expectation values, hence yielding zero total magnetization. The different plots resemble different temperatures: for $T_{R}=1K$ blue, for $T_{R}=2K$ red and for $T_{R}=3K$ green.}
	\label{sp2}
\end{figure}

From the previous analysis we can see that for both magnetization protocols PMD and AMD respectively, the local spin expectation value is finite, and hence one must expect the total magnetization to not vanish.  
\begin{figure}[!ht]
	\centering
	\includegraphics[width=0.45\textwidth]{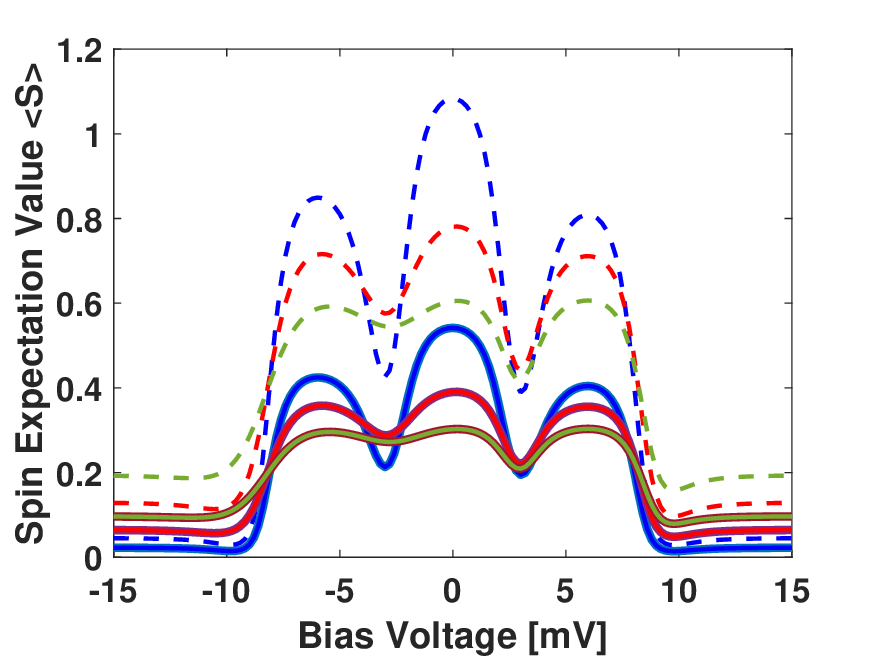}
	\caption{Total Molecular Magnetization Under PMD: here the $\left\langle S^{(z)}_{a}+S^{(z)}_{b}\right\rangle$ is ploted in dashed lines and in continues lines $\left\langle S^{(z)}_{a}\right\rangle=\left\langle S^{(z)}_{b}\right\rangle$ is being plotted. The convenctions are: blue for $1K$, red for $2K$ and green for $3K$.}
	\label{sp1}
\end{figure}
For AMD, in fig. \ref{beff} is shown that the asymptotic divergence with respect to the zero external magnetization condition of the effective local magnetization does not mean that the net magnetization is zero, mainly due to the spin asymmetry in the junction, possibly except for the zero external magnetization condition.
\\
To just illustrate the effect of PMD on the total molecular magnetization we evaluate the local expectation values on each spin of the dimer and add them to gain understanding on how the total magnetization will behave, and clearly it does as a ferromagnet as shown in fig. \ref{sp1}, at least in the temperature range $1K-10K$.
In fig. \ref{sp2}, we illustrate the spin expectation value of each spin unit under the AMD protocol, showing an asymptotic behavior that yields zero total moment, resembling the argument that was originally given in \cite{Saygun2016a}.
Basic quantities for assessing quantum transport can be evaluated from the Jauho-Meir-Wingren formalism \cite{Jauho1994}. Here in use an alternative version of this formalism derived in \cite{VasquezJaramillo2018b}, where the charge current injected from any of the metallic leads in to the molecule is given by:
\begin{align}
J_{\alpha}^{(c)}(V,\Delta T)= \frac{ie}{\hbar}\int_{-\infty}^{+\infty}\frac{d\epsilon}{2\pi}\Gamma_{nm\sigma\sigma}^{(\alpha)}(\epsilon)&\Big(G_{mn,\sigma\sigma}^{>}(\epsilon)f_{\sigma}(\epsilon)
\nonumber
\\
&+G^{<}_{mn,\sigma\sigma}(\epsilon)f_{\sigma}(-\epsilon)\Big),
\label{qt1}
\end{align}
where $\Gamma_{nm\sigma\sigma}^{(\alpha)}(\epsilon)$ is the molecule-lead coupling matrix element in the site and spin space, $G_{mn,\sigma\sigma}^{</>}(\epsilon)$ is the lesser/greater Green's function depending on the spin dependent energy band in the continuous limit $\epsilon$, and $f_{\sigma}(\epsilon)$ stands for the spin-resolved Fermi-Dirac distribution and this expression is completely consistent with reference \cite{Galperin2006}. Note the dependence on the bias voltage $V$ and the Temperature difference $\Delta T$ of the charge current due to the Fermi-Dirac distribution. Here, the effect of the spin expectation values evaluated from expressions \ref{expec12a}, comes in the retarded Green's functions $G_{mn,\sigma\sigma}^{R}(\epsilon)$ which is related to the lesser/greater Green's functions $G_{mn,\sigma\sigma}^{</>}(\epsilon)$ through the Keldysh equation given by expression \ref{keldysh_eq}. See supplementary material for details.

\begin{figure}[!ht]
	\centering
	\includegraphics[width=0.45\textwidth]{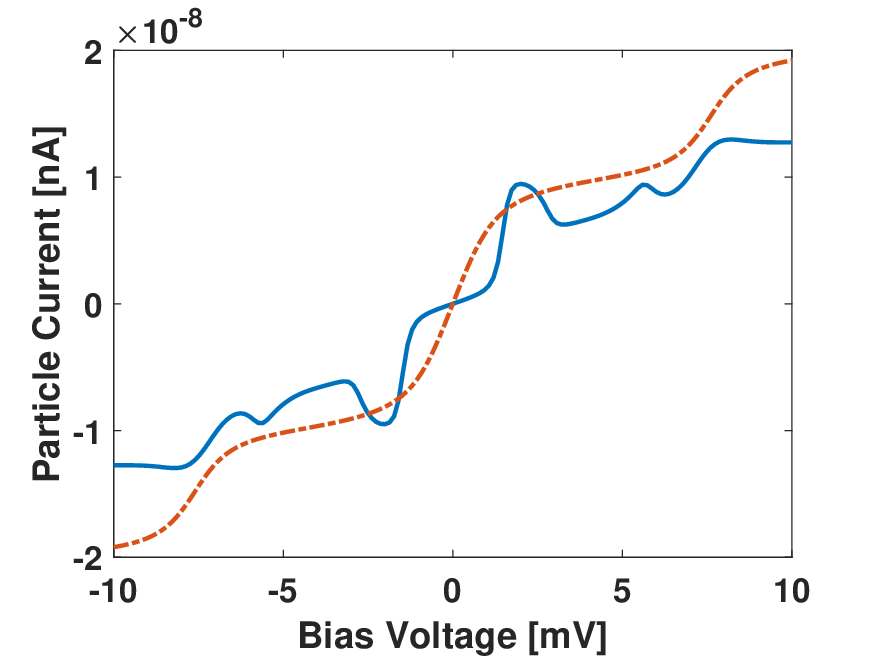}
	\caption{Illustration of the particle current for finite spin expectation value in each individual spin unit (blue, continuous line) and of the particle current with no spin moments (bare current) (red, dashed line).}
	\label{transport1}
\end{figure}

\begin{figure}[!ht]
	\centering
	\includegraphics[width=0.45\textwidth]{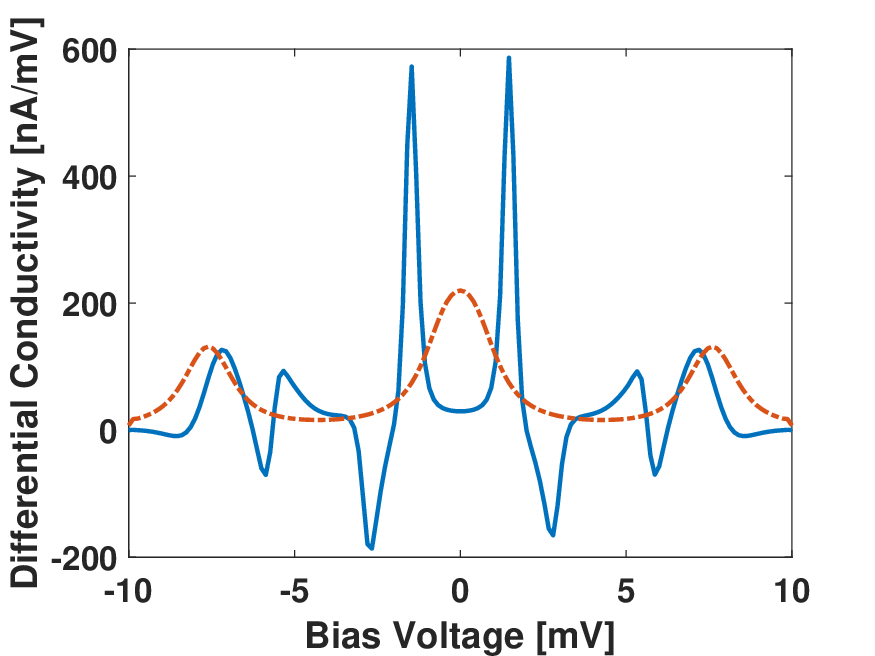}
	\caption{Differential conductivity of a magnetic tunneling junction: Red dotted line for the case in which no symmetry breaking interaction is considered. Continuous blue line for the case of a magnetized tunneling junction, here specifically with the PMD protocol}
	\label{transport2}
\end{figure}

In figure \ref{transport1} the current defined in expression \ref{qt1} is being plotted against bias voltage for $\Delta T=2.9K$. In the absence of symmetry breaking interactions (ferromagnetic contacts and/or external magnetic induction fields), the charge current is plotted as a dotted red line. For a finite external magnetic induction $B=4T$ under a PMD protocol and in the presence of ferromagnetism in both metallic leads spin polarized at $50\%$ with their magnetizations oriented along the $z-$ direction for both of them, the current is plotted as a continuous blue line. As compare to the results published in references \cite{Saygun2016a,Vasquez-Jaramillo2017}, note that the zero bias behavior shown in figure \ref{transport1} is consistent with the one reported in these references, in the sense that the conductivity around zero bias in the absence of symmetry breaking interactions is relatively large, and when the spin expectation values are taken into account the spin dimer is configured in a spin singlet (see figures \ref{exchange}), which localizes the electronic wave functions hence achieving low differential conductivity (change in current with respect to a change in bias voltage), exactly as shown in figure \ref{transport2}.

Note that for the case of a four-fold degeneracy of the spin Hamiltonian eigenvalues the differential conductivity is again zero due to the fact that the charge current becomes stationary with voltage (this will change when considering higher order contributions to the quantum transport with respect to the tunneling parameter).

\section{Discussion and Conclusions}
The work presented here, is intended to contribute in three different aspects to the problem of magnetized spin molecular junctions:
\begin{itemize}
    \item Demonstrate, that contrary to what was reported in references \cite{Saygun2016a,Vasquez-Jaramillo2017}, completely paramagnetic molecular junctions do not exhibit individual spin magnetic moments, and hence, no spin structure signatures are observed in quantum transport quantities.
    \item Investigate the conditions under which the spin asymmetry can be induced in the molecule, that is, what are the circumstances under which external magnetization is transferred into the molecule.
    \item Propose different mechanisms that allow for spin asymmetry to be transferred into the molecule. In the present work, we have proposed and investigated two magnetization protocols at the molecule level: Parallel Magnetic Driving (PMD) and Anti-Parallel Magnetic Driving (AMD).
\end{itemize}

With respect to the impossibility of a paramagnetic exchange couple dimer to exhibit finite moments for individual spin units, we have analytically demonstrated a well known result of statistical mechanics in expression \ref{expec12a}, which is, that in the absence of a magnetic induction field, each individual unit has zero moment, as it can be appreciated from figures \ref{moment1} and \ref{moment2}. This perspective, can be extended to more than two spins, for instance, a trimer of spins as considered in reference \cite{FernandezSanchez2021} (See the results for a spin trimer in the supporting information, where the each individual spin moment is plotted as a function of the exchange couplings giving finite total moment). In the case which is particularly consider in this work, we identify two important mechanisms to break a symmetry that will make the spin moments in each unit different than zero, being one of them transferring the spin asymmetry from the ferromagnetic leads into the molecule, which becomes a nonequilibrium effect driven by the bias voltage as shown in figure \ref{beff}, and the second mechanism, which is the more conventional one, being the application of an external magnetic field. Breaking this symmetries will then produced finite spin moments which is then projected into the retarded Green's function (See Supporting Information). In this particular way, the spin moment information in included in the nonequilibrium Green's functions which are defined in terms of the retarded and advanced Green's functions through the Keldysh equation given by expression \ref{keldysh_eq}, and of particular interest is the signatures of the spin structure in the transport, which become then spin dependent when the spin moments are finite at low temperatures and finite effective induction fields. In figures \ref{transport1} and \ref{transport2} we can appreciate the difference when spin structure is reflected in quantum transport quantities (continuous blue line) and when is not (dotted red line).  With respect to investigating the conditions under which the magnetization can be transfer into the molecule, we have shown in figure \ref{dimer} that even in the absence of external magnetic induction fields, there is a finite effective magnetic induction field which depends on the applied bias voltage in the junction. We also show that this effective field is anti-symmetric around zero bias through the junction as expected since it resembles a Zeeman energy derived from nonequilibrium field theory in reference \cite{VasquezJaramillo2018b} as a first order correction to the spin action given by expression \ref{eff1}. We as well show that for large absolute value of the bias voltage, the magnitude of the effective magnetic induction field is independent of the value of the external magnetic induction. With regards to the magnetization protocols, that is PMD and AMD, we have shown as well that it influences quite notoriously the nonequilibrium induced symmetric exchange coupling and the nonequilibrium induced anti-symmetric chiral anisotropic exchange coupling. For instance, in figure \ref{exchange} we can appreciate that external magnetic induction fields applied to the molecule using the PMD protocol, induced an asymmetry around zero bias and as well produces non-four-fold degenerate spin states at large bias voltages, which was the case in the absence of external induction fields. With respect to the spin excitation spectra, it was clearly observed that in the absence of external and induced magnetization, there is a triplet degeneracy that makes three of the our quantum states indistinguishable. To lift this degeneracy, symmetry breaking is required, and here we use external magnetization and induced magnetization by proximity. The latter, lifts the triplet degeneracy, hence, making all states distinguishable and accesable electrically. This particular observation, notoriously contrasts with the simulations reported in \cite{Saygun2016a,Vasquez-Jaramillo2017}, as in this previous cases,  the quantum transport at large bias voltages, was dominated by the fact that the spin states were four-fold degenerate and the molecular junction will behave like a Landauer type transport channel. In the case presented here, as appreciated from figure \ref{transport2}, there is a variety of trends in the differential conductivity that were not presented in the previous references which is purely due to the magnetization protocol. For the nonequilibrium induced anti-symmetric chiral anisotropic exchange interaction when the junction is magnetically driven using the PMD protocol, as expected, the applied magnetic induction competes with the chirality, and in fact, large magnetic induction reverses the chirality as it can be seen from figure \ref{DM}. For the case of AMD, in figure \ref{exchange1} it can be observed, that in the nonequilibrium induced symmetric and isotropic exchange coupling, there is a Kondo like zero bias anomaly, though, for the parameter space considered in this simulation, there is no shift in the nature of the exchange coupling between spins (For instance in reference \cite{VasquezJaramillo2019}, we consider the effect of the Aharonov-Bohm phase in the exchange coupling which is, that the nature of the alignment between spins shifts from ferromagnetic to anti-ferromagnetic not only as a function of the phase but also as a function of the induction field).
\\
\\
The work presented in this article well establishes the correct methodology to map the spin structure of a molecule into quantum transport measurements, and in that way setting the bases for the correct methodology to implement spin resolved scanning tunneling spectroscopy SR-STS on molecules with spin degrees of freedom. This work also establishes two novel protocols to drive magnetic molecular junctions and deeply accounts for the relevant features and trends in the molecular magnetism that rely upon each of the protocols.

\begin{acknowledgements}
We would like to acknowledge the contribution of Jhoan Alexis Fernandez Sanchez and Luis Alejandro Sierra Ossa for critical reading of the manuscript and analysis of the spin moments for the case of a spin trimer. We thank Prof. Adrian Kantian from Uppsala University for the useful insight regarding the problem of the magnetization of magnetic molecules. We also acknowledge Prof. Jonas Fransson from Uppsala University, who kindly shared his ideas on spin resolved nonequilibrium physics with Juan David during his Ph.D, and whose work we now extend. We acknowledge funding from Minciencias (Ministry of Science, Technology and Innovation - Colombia) and From Universidad del Valle through the grant for the project with reference CI 71261 corresponding to the call by Minciencias No. 649 from 2019 for postdoctoral research scholars under the agreement number 80740-618-2020.
\end{acknowledgements}

%%%%%%%%%%%%%%%%%%%%%%%%%%%%%%%%%%%%%%%%%%

\appendix

\section{Retarded Green's Function in the Presence of Spin}

From reference \cite{Vasquez-Jaramillo2017} and in reference \cite{VasquezJaramillo2018b}, the retarded Green's function for a spin dimer can be written as follows:
\begin{widetext}
\begin{equation}
G(\omega)=\left[\begin{array}{cc}\hbar\omega-\epsilon_{a\sigma}-\mathbb{J}_{a}\langle\bm{S}_{a}^{(z)}\rangle+\frac{i\Gamma_{a\sigma}}{2}&-\gamma_{ab}\\
-\gamma_{ab}&\hbar\omega-\epsilon_{b\sigma}-\mathbb{J}_{b}\langle\bm{S}_{b}^{(z)}\rangle+\frac{i\Gamma_{b\sigma}}{2}\end{array}\right]^{-1}
\end{equation}
\end{widetext}
With:
\begin{align}
\Gamma_{a\sigma}=\Gamma_{a\sigma}^{(L)}+\Gamma_{a\sigma}^{(R)},~~~~~~~
\Gamma_{b\sigma}=\Gamma_{b\sigma}^{(L)}+\Gamma_{b\sigma}^{(R)}
\end{align}
Here the parameters are as defined previously in figure \ref{dimer}. Note that the spin moments renormalizes each energy level, and hence, affects the number of electrons allow through each transport channel. This is the very cornerstone of the spin structure signature in the quantum transport quantities.

\section{Evaluation of Matrix Elements For the Spin Moments in the Singlet-Triplet Basis}
\begin{widetext}
\begin{align}
	\braket{0|\bm{S}^{(z)}_{a}|0}&=\frac{1}{2}\left(\braket{\uparrow\downarrow|\bm{S}^{(z)}_{a}|\uparrow\downarrow}-
	\braket{\uparrow\downarrow|\bm{S}^{(z)}_{a}|\downarrow\uparrow}
	 -  \braket{\downarrow\uparrow|\bm{S}^{(z)}_{a}|\downarrow\uparrow}
	+\braket{\downarrow\uparrow|\bm{S}^{(z)}_{a}|\downarrow\uparrow}\right)=0,
	\label{expec4}
	\\
	\braket{1|\bm{S}^{(z)}_{a}|1}&=
	\frac{1}{2}\left(\braket{\uparrow\downarrow|\bm{S}^{(z)}_{a}|\uparrow\downarrow}+
	\braket{\uparrow\downarrow|\bm{S}^{(z)}_{a}|\downarrow\uparrow}
	+\braket{\downarrow\uparrow|\bm{S}^{(z)}_{a}|\downarrow\uparrow}
	+\braket{\downarrow\uparrow|\bm{S}^{(z)}_{a}|\downarrow\uparrow}\right)=0
	\label{expec5}
	\\
	\braket{2|\bm{S}^{(z)}_{a}|2}&=\braket{\uparrow\uparrow|\bm{S}^{(z)}_{a}|\uparrow\uparrow}=\frac{\hbar}{2},
	\label{expec6}
	\\
	\braket{3|\bm{S}^{(z)}_{a}|3}&=\braket{\downarrow\downarrow|\bm{S}^{(z)}_{a}|\downarrow\downarrow}=-\frac{\hbar}{2},
	\label{expec7}
\end{align}
\end{widetext}

\section{Evaluation of the Expectation Value of the Spin Units and the Partition Function for the case of a Molecular Paramagnetic Dimer}
    Now let's evaluate the average $j-th$ magnetic moment from:
    
    \begin{equation}
	\langle\bf{S_{j}}^{(z)}\rangle=\frac{1}{\mathcal{Z}}\textit{Tr}\left[e^{-\beta\bf{\mathcal{H}}_{spin}}\bf{S_{j}}^{(z)}\right],
	\end{equation}
	where $\mathcal{Z}$ is the partition function and it can be evaluated as follows:

 \begin{align}	\mathcal{Z}&=\textit{Tr}\left[e^{\beta\bf{\mathcal{H}}_{spin}}\right]=\sum_{n=0}^{3}\braket{n|[e^{\beta\bf{\mathcal{H}}_{spin}}|n}
    =\sum_{n=0}^{3}e^{-\beta\epsilon_{n}}
    \nonumber\\
    &=e^{-\beta\left(\frac{\mathcal{J}}{4}\right)}+e^{-\beta\left(\frac{\mathcal{J}}{4}+\frac{\Delta B_{z}}{2}\right)}+e^{-\beta\left(\frac{\mathcal{J}}{4}-\frac{\Delta B_{z}}{2}\right)}+e^{\beta\left(\frac{3\mathcal{J}}{4}\right)}
	\nonumber\\
    &=e^{-\frac{\beta\mathcal{J}}{4}}+e^{-\frac{\beta\mathcal{J}}{4}}
	e^{-\frac{\beta\Delta B_{z}}{2}}+e^{-\frac{\beta\mathcal{J}}{4}}e^{\frac{\beta\Delta B_{z}}{2}}
	+e^{\frac{3\beta\mathcal{J}}{4}}
	\nonumber\\
    &=e^{-\frac{\beta\mathcal{J}}{4}}+e^{-\frac{\beta\mathcal{J}}{4}}\left(
	e^{\frac{\beta\Delta B_{z}}{2}}+e^{-\frac{\beta\Delta B_{z}}{2}}\right)+
	e^{\frac{3\beta\mathcal{J}}{4}}
	\nonumber\\
    &=e^{-\frac{\beta\mathcal{J}}{4}}\left(1+2\cosh\left[\frac{\beta\Delta B_{z}}{2}\right]+e^{\beta\mathcal{J}}\right),
\end{align}

giving:
\begin{align}
\langle\bf{S_{j}}^{(z)}\rangle &=\frac{1}{\mathcal{Z}}\textit{Tr}\left[e^{-\beta\bf{\mathcal{H}}_{spin}}\bf{S_{j}}^{(z)}\right]
    \nonumber
    \\
    &=\frac{\textit{Tr}\left[e^{-\beta\bf{\mathcal{H}}_{spin}}\bf{S_{j}}^{(z)}\right]}{e^{-\frac{\beta\mathcal{J}}{4}}\left(1+2\cosh\left[\frac{\beta\Delta B_{z}}{2}\right]+e^{\beta\mathcal{J}}\right)}
\end{align}

Now, let's see how to evaluate $\textit{Tr}\left[e^{-\beta\bf{\mathcal{H}}_{spin}}\bf{S_{j}}^{(z)}\right]$:
\begin{align}
		\textit{Tr}&\left[e^{-\beta\bf{\mathcal{H}}_{spin}}\bf{S_{1}}^{(z)}\right]=
		\sum_{n=0}^{3}\braket{n|e^{-\beta\bf{\mathcal{H}}_{spin}}\bf{S_{1}}^{(z)}|n}=\frac{\hbar}{2}e^{-\beta\epsilon_{3}}-\frac{\hbar}{2}e^{-\beta\epsilon_{2}}
		\nonumber
		\\
		&=\frac{\hbar}{2}\left(e^{-\beta\left(\frac{\mathcal{J}}{4}-\frac{\Delta B_{z}}{2}\right)}-e^{-\beta\left(\frac{\mathcal{J}}{4}\frac{\Delta B_{z}}{2}\right)}\right)
        =\frac{e^{-\frac{\beta\mathcal{J}}{4}}\hbar}{2}
		\left(e^{\frac{\beta\Delta B_{z}}{2}}-e^{-\frac{\beta\Delta B_{z}}{2}}\right)
		\nonumber
		\\
		&=\hbar e^{-\frac{\beta\mathcal{J}}{4}}\sinh\left[\frac{\beta \Delta B_{z}}{2}\right].
\end{align}
and replacing in the expression for the spin expectation value we arrive at expression \ref{expec12a}.

\section{Solving Eigenvalue Problems}
For the PMD protocol:
\begin{widetext}
\begin{align}
	|\mathcal{\bm{H}}_{spin}(\xi=1)-\mathcal{E}\mathbb{I}|&=
	\left|\begin{array}{cccc}\mathcal{J}_{ab}-\Delta B^{(z)}_{eff}-\mathcal{E}&0&0&0
		\\
		0&-\mathcal{J}_{ab}-\mathcal{E}&2\mathcal{J}_{ab}&0
		\\
		0&2\mathcal{J}_{ab}&-\mathcal{J}_{ab}-\mathcal{E}&0
		\\
		0&0&0&\mathcal{J}_{ab}+\Delta B^{(z)}_{eff}-\mathcal{E}
	\end{array}\right|,
	\nonumber
	\\
	&=\left(\mathcal{J}_{ab}-\Delta B^{(z)}_{eff}-\mathcal{E}\right)\left[\left(\mathcal{J}_{ab}+\mathcal{E}\right)^{2}\left(\mathcal{J}_{ab}+\Delta B^{(z)}_{eff}-\mathcal{E}\right)-(2\mathcal{J}_{ab})^{2}\left(\mathcal{J}_{ab}+\Delta B^{(z)}_{eff}-\mathcal{E}\right)\right],
	\nonumber
	\\
	&=\left(\mathcal{J}_{ab}-\Delta B^{(z)}_{eff}-\mathcal{E}\right)\left(\mathcal{J}_{ab}+\Delta B^{(z)}_{eff}-\mathcal{E}\right)\left[\left(\mathcal{J}_{ab}+\mathcal{E}\right)^{2}-4\mathcal{J}_{ab}^{2}\right],
	\nonumber
	\\
	&=\left(\mathcal{J}_{ab}-\Delta B^{(z)}_{eff}-\mathcal{E}\right)\left(\mathcal{J}_{ab}+\Delta B^{(z)}_{eff}-\mathcal{E}\right)\left(-\mathcal{J}_{ab}+\mathcal{E}\right)
	\left(3\mathcal{J}_{ab}+\mathcal{E}\right)=0,
\end{align}
Giving:
\begin{align}
\mathcal{E}_{0}&=-3\mathcal{J}_{ab};~~~~~~~~~~~~~~~~\mathcal{E}_{1}=\mathcal{J}_{ab}.
    \nonumber \\
    &\mathcal{E}_{2}=\mathcal{J}_{ab}+\Delta B^{(z)}_{eff};~~~~\mathcal{E}_{3}=\mathcal{J}_{ab}-\Delta B^{(z)}_{eff}
\end{align}

For the AMD protocol, we have instead:

\begin{align}
	|\mathcal{\bm{H}}_{spin}(\xi=-1)-\mathcal{E}\mathbb{I}|&=\left|\begin{array}{cccc}\mathcal{J}_{ab}-\lambda&0&0&0
		\\
		0&-\mathcal{J}_{ab}-\Delta B^{(z)}_{eff}-\lambda&2\mathcal{J}_{ab}&0
		\\
		0&2\mathcal{J}_{ab}&-\mathcal{J}_{ab}+\Delta B^{(z)}_{eff}-\lambda&0
		\\
		0&0&0&\mathcal{J}_{ab}-\lambda
	\end{array}\right|,
	\nonumber
	\\
	&=\left(\mathcal{J}_{ab}-\lambda\right)\left[\left(-\mathcal{J}_{ab}-\Delta B^{(z)}_{eff}-\lambda\right)\left(-\mathcal{J}_{ab}+\Delta B^{(z)}_{eff}-\lambda\right)\left(\mathcal{J}_{ab}-\lambda\right)-(2\mathcal{J}_{ab})^{2}
	\left(\mathcal{J}_{ab}-\lambda\right)\right],
	\nonumber
	\\
	&=\left(\mathcal{J}_{ab}-\lambda\right)^{2}
	\left[\left(\mathcal{J}_{ab}+\Delta B^{(z)}_{eff}+\lambda\right)\left(\mathcal{J}_{ab}-\Delta B^{(z)}_{eff}+\lambda\right)-(2\mathcal{J}_{ab})^{2}\right],
	\nonumber
	\\
	&=\left(\mathcal{J}_{ab}-\lambda\right)^{2}
	\left[\left(\mathcal{J}_{ab}+\lambda\right)^{2}-\left(\Delta B^{(z)}_{eff}\right)^{2}-(2\mathcal{J}_{ab})^{2}\right]
	\nonumber
	\\
	&=\left(\mathcal{J}_{ab}-\lambda\right)^{2}
	\left[\lambda^{2}+2\mathcal{J}_{ab}\lambda-\left(\Delta B^{(z)}_{eff}\right)^{2}-3\mathcal{J}^{2}_{ab}\right]=0,
% 	\label{expec23}
\end{align}
\end{widetext}

Giving the eigenvalues $\mathcal{E}_{0}$ and $\mathcal{E}_{1}$ as follows:
$$
\mathcal{E}_{0}=\mathcal{E}_{1}=\mathcal{J}_{ab}
$$
and for eigenvalues $\mathcal{E}_{2}$ and $\mathcal{E}_{3}$ these are given by expression \ref{expec26} and an approximate ones in expressions \ref{expec28} and \ref{expec29}.
\\
\newpage

\section{Extension to the Case of Three Spins}
The case of a spin trimer is been considered in reference \cite{FernandezSanchez2021}, where we can see the effect of the temperature and the external magnetization in the spin moment of individual spin units and and in the total spin moment. Note that for the spin trimer all spin units are exchange coupled.
\begin{figure}[!ht]
	\centering
	\includegraphics[width=0.45\textwidth]{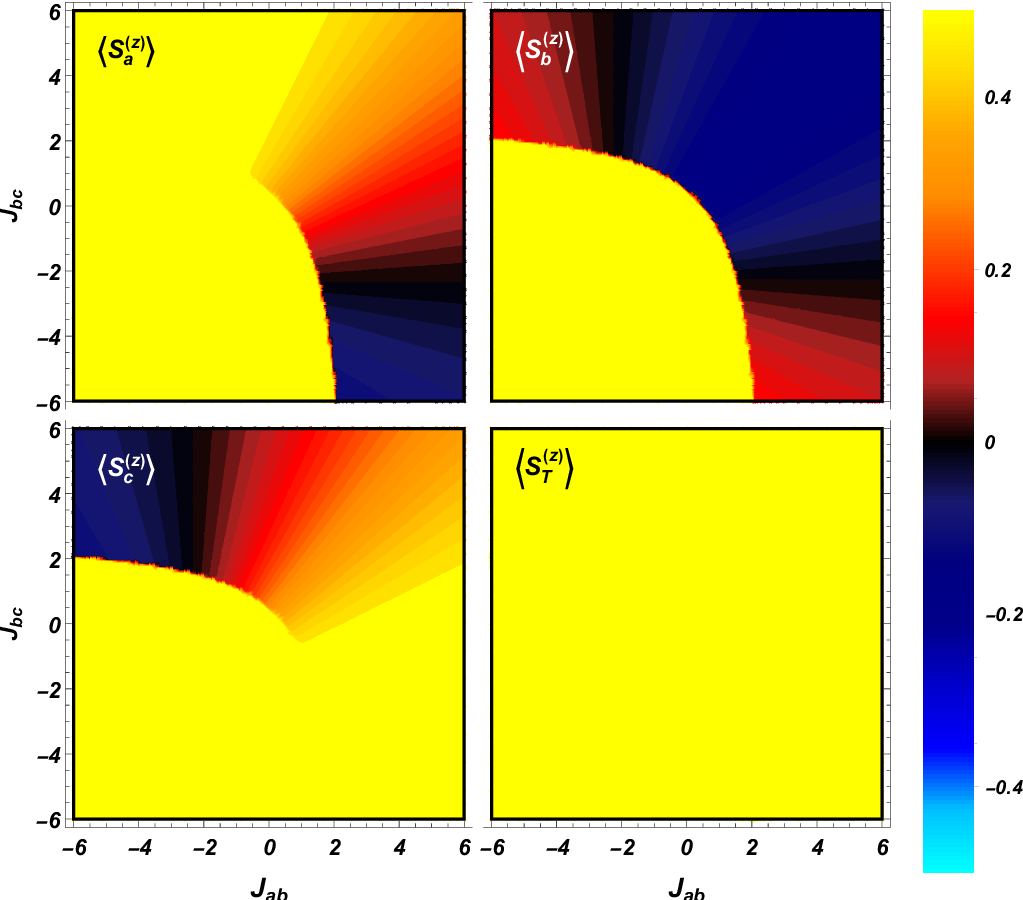}
	\caption{Spin moment on each of the spin units of the spin trimer and the total spin moment of the molecule in the case of PMD between each spin unit. The temperature consider here as $T=0.1K$}
	\label{al1}
\end{figure}
\newpage
\begin{figure}[!ht]
	\centering
	\includegraphics[width=0.45\textwidth]{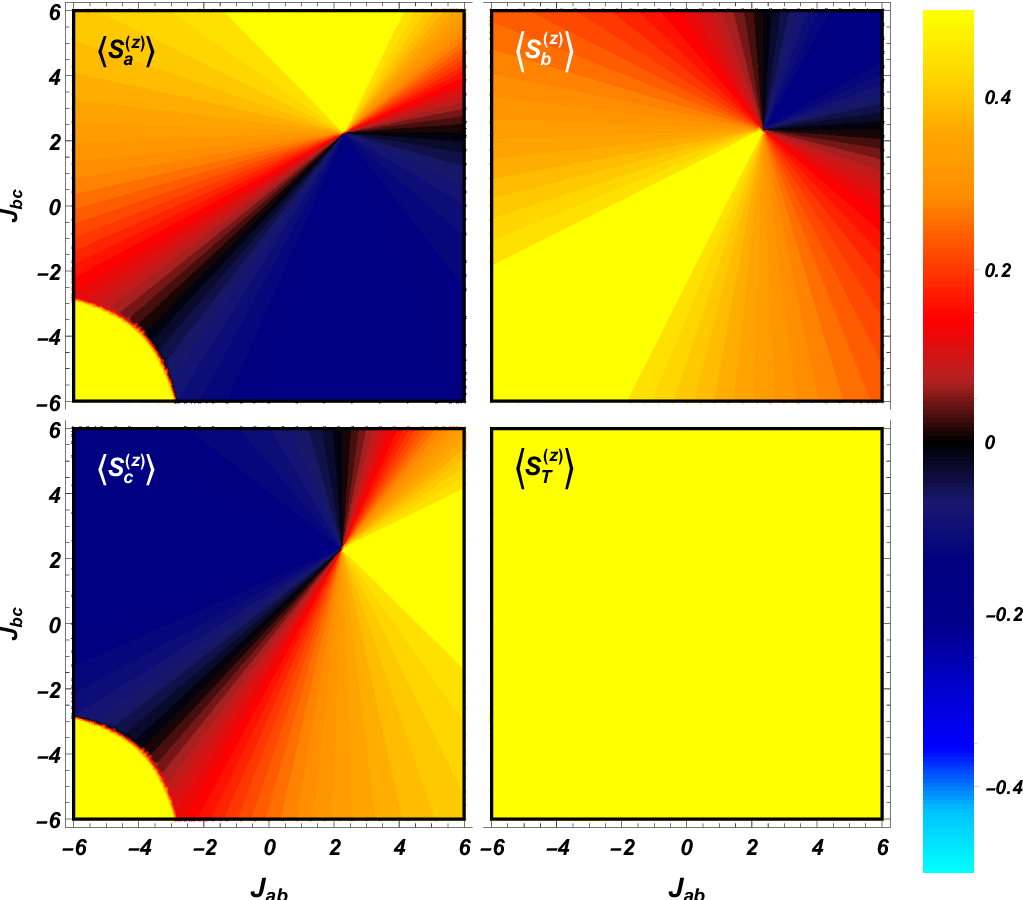}
	\caption{Spin moment on each of the spin units of the spin trimer and the total spin moment of the molecule in the case of PMD between each spin unit. The temperature consider here as $T=0.1K$ and we fix the exchange coupling of two units to $J=2.3meV$}
	\label{al2}
\end{figure}

%%%%%%%%%%%%%%%%%%%%%%%%%%%%%%%%%%%%%%%%%%%%%%%%%%%%%%%%%%%%%%%%%%%%%%%%%%%%%%%%%%%%%%%%%%%%%%%%%%%%%%%%%%%%%%%%%%%%%%%%%%%%%%%%%%%%%%%%%%%%%%%%%%%%%%%%%%%%%%%%%%%%%%%%%%%%%%%%%%%%%%%%%%%%%%%%%%%%%%%%%%%%%%%%%%%%%%%%%%%%%%%%%%%%%%%%%%%%%%%%%%%%%%%%%%%%%%%%%%%%%%%%%%%%%%%%%%%%%%%%%%%%%%%%%%%%%%%%%%%%%%%%%%%%%%%%%%%%%%%%%%%%%%%%%%%%%%%%%%%%%%%%%%%%%%%%%%%%%%%%%%%%%%%%%%%%%%%%%%%%%%%%%%%%%%%%%%%%%%%%
\bibliography{dimer,dimer1,dimer2}% Produces the bibliography via BibTeX.

\end{document}